\documentclass[pdflatex,sn-mathphys-num]{sn-jnl}

\usepackage{graphicx}%
\usepackage{multirow}%
\usepackage{amsmath,amssymb,amsfonts}%
\usepackage{amsthm}%
\usepackage{mathrsfs}%
\usepackage[title]{appendix}%
\usepackage{xcolor}%
\usepackage{textcomp}%
\usepackage{manyfoot}%
\usepackage{booktabs}%
\usepackage{algorithm}%
\usepackage{algorithmicx}%
\usepackage{algpseudocode}%
\usepackage{listings}%

\usepackage{placeins}

\usepackage{subcaption}

\usepackage{xspace} 
\usepackage{bigstrut}
\usepackage{booktabs}

\theoremstyle{thmstyleone}%
\theoremstyle{thmstyletwo}%

\theoremstyle{thmstylethree}%

\newcommand {\un}[1]{\boldsymbol{#1}}

\usepackage{xspace} 

\providecommand{\TA}{\texttt{tas}\xspace}

\providecommand{\AC}{\texttt{ACCESS-CM2}\xspace}

\providecommand{\CE}{\texttt{CESM2}\xspace}
\providecommand{\EC}{\texttt{EC-Earth3}\xspace}
\providecommand{\MR}{\texttt{MRI-ESM2-0}\xspace}

\providecommand{\UK}{\texttt{UKESM1-0-LL}\xspace}

\providecommand{\SL}{\texttt{SSP126}\xspace}
\providecommand{\SM}{\texttt{SSP245}\xspace}
\providecommand{\SH}{\texttt{SSP585}\xspace}

\providecommand{\DAN}{{Antarctic}\xspace}
\providecommand{\DDA}{{Dasht-e Lut}\xspace}
\providecommand{\DMO}{{Mojave}\xspace}
\providecommand{\DSA}{{Sahara}\xspace}
\providecommand{\DSI}{{Simpson}\xspace}
\providecommand{\DUK}{{UK}\xspace}

\newcommand{\ed}[1]{\textcolor{black}{#1}}
\definecolor{lliw-ed}{rgb}{0, 0, 0}

\newcommand\Tstrut{\rule{0pt}{2.6ex}} 

\begin{document}

\title[Model selection in extreme value regression]{{Model selection in extreme value regression for estimation of changes in return values over time applied to extreme annual regional CMIP6 desert temperatures}}


\author[1]{\fnm{Callum} \sur{Leach}}\email{c.r.leach@lancaster.ac.uk}

\author[2]{\fnm{Kevin} \sur{Ewans}}\email{kevin.ewans@mrl.nz}

\author*[3]{\fnm{Philip} \sur{Jonathan}}\email{p.jonathan@lancaster.ac.uk}

\affil[1]{\orgname{Lancaster University}, \orgaddress{\city{Lancaster}, \postcode{LA1 4YQ}, \country{UK}}}

\affil[2]{\orgname{MetOcean Research Ltd}, \orgaddress{\city{New Plymouth}, \postcode{4310}, \country{New Zealand}}}

\affil[3]{\orgname{Lancaster University}, \orgaddress{\city{Lancaster}, \postcode{LA1 4YF}, \country{UK}}}


\abstract{Estimating changes in extremes quantiles from environmental processes which are non-stationary in time from small samples is challenging, since it is difficult to characterise the nature of tail {non-stationarity} adequately, and hence to estimate extreme quantiles in time. Using annual maxima and minima of {near-surface temperature (\TA)} from {Coupled Model Intercomparison Project Phase 6 (CMIP6) output} for a number of the Earth's desert regions as illustration, we use generalised extreme value (GEV) regression to characterise changes in extreme quantiles over the next century. We consider a set of candidate models with different parametric forms for the variation of GEV parameters with time, estimating parameters using Bayesian inference. We seek to select the optimal candidate model using one of a number of model selection criteria, including the Akaike, Bayesian, divergence and widely-applicable ``information criteria''. However, in an extreme value setting, the performance of different model selection criteria is unreliable. We therefore undertake a simulation study using ground truth models generating data qualitatively similar to annual temperature extrema, to assess the relative performance of the criteria to minimise error in prediction of {change $\Delta Q$ in the 100-year return value of \TA over the period (2015,2125)}. In our application, the Bayesian information criterion (BIC) provides best performance, clearly out-performing the divergence and widely-applicable information criteria in particular. We compare BIC model selection with a stacked Bayesian model average. Using BIC-selected GEV regression models, we estimate joint posterior distributions of $\Delta Q$, coupled over three climate scenarios, for different combinations of desert region, global climate model and climate ensemble. Aggregating posterior estimates over {climate models and ensembles}, we find evidence for significant increases in $\Delta Q$ for regional annual maxima under the stronger forcing scenarios for all desert regions. Similar but weaker and less significant trends are observed for regional annual minima. {There is evidence that extreme minima are warming more rapidly than extreme maxima in Antarctica, whereas most hot desert regions exhibit the opposite effect. \\ \vspace{1pt}\\Ancillary files: supplementary material is provided as an ancillary file with this submission.}} 




\maketitle

\clearpage
\section{Introduction} \label{Sct:Int}
%
{Our planet's climate is changing due to human activity. Projections under specified future emissions scenarios using global climate models (GCMs) allow assessment of risk to society, and the efficacy of remedial actions. Large uncertainties in projections remain, especially at local scales (see e.g. \cite{SmpEA25}). Some of the greatest societal impacts are caused by changes in mean characteristics of variables quantifying the state of the Earth's atmosphere and oceans. The fidelity with which these changes are encoded in GCMs can be assessed by comparing climate model output with observation. Other important impacts are caused by extremes, including heatwaves and storms. It is therefore critical also to understand the characteristics of projections of extremes from GCMs. Near-surface atmospheric temperature (\TA) is a critical output of global climate models; it is widely observed, it is the variable most commonly used to define global warming. It is typically projected to increase under all emission scenarios, and it is a key parameter in physical processes. In particular, the Sixth Assessment Report of the Intergovernmental Panel on Climate Change (\cite{IPCC-2023SPM}) concludes with high to very high confidence that continued increases in global \TA will lead to systematic intensification of climate impacts, with severities strongly dependent on the magnitude of warming. For example, it is virtually certain that the frequency and intensity of hot extremes, including heatwaves, will increase as global mean near-surface temperature rises.}

Non-stationary extreme value analysis is widely applied in environmental science and engineering. Estimating the nature of the non-stationarity however can be problematic, especially when sample size is small. In elementary terms, we seek to estimate a statistical model for the probability distribution which generated the sample data, the functional form of which is dependent on a number of parameters (e.g. the location, scale and shape of a generalised extreme value (GEV) distribution). In a non-stationary model, model parameters themselves vary with covariates. Often, the form of the functional relationship for a model parameter with covariates is assumed to take a parametric form; e.g. a model parameter might be linear or quadratic in one or more covariates. Given a set of such ``parametric regression'' candidate models, we can use model selection to identify the parametric form for model parameter variation with covariates which is most consistent with the sample data. Unfortunately, as discussed in the next paragraph, reliable model selection for non-stationary extreme value analysis is difficult to achieve. An alternative approach to parametric regression is ``semi-parametric'' (or ``non-parametric'') regression, in which variation of model parameters with covariates is described using more flexible representations (such as splines, Gaussian processes or neural networks). Now there is no need for model selection as such: the challenge instead is to restrict the flexibility of covariate representation so that the model gives optimal performance (see e.g. \cite{ZnnEA19a}). 

For parametric regression, there are numerous approaches to model selection. If the objective is minimisation of generalisation error, then maximisation of model out-of-sample predictive performance is an appropriate strategy (see e.g. \cite{VhtOjn12}, \cite{GlmEA13}). This assessment tends to involve either repeated model fitting (e.g. in a cross-validation scheme) or model fitting to just a subset of the sample (e.g. if the sample has been partitioned into training and test subsets). An alternative approach, popular amongst practitioners in particular, is the use of cost-complexity scores or ``information criteria'' to rank candidate models, selecting the model with optimal score. This approach is relatively straightforward, requiring just one model fit to the sample. An information criterion typically penalises a model in terms of both its lack of fit to the sample, and its complexity; a good low score requires relatively good fit and relative parsimony. Different information criteria quantify lack of fit and model complexity differently, motivated by theoretical and practical considerations; \cite{GlmEA13} \cite{GlmEA14}, \cite{Kim17} and \cite{ZhnEA23} provide overviews. Here we consider the Akaike Information Criterion (AIC, \cite{Akk74}), the Bayesian Information Criterion (BIC, \cite{Sch78}), the Divergence Information Criterion (DIC, \cite{SpgEA02}, \cite{GlmEA13}) and the Widely Applicable Information Criterion (WAIC, \cite{Wtn13},\cite{VhtEA17}) for model selection. Supplementary material Section~SM1 provides further discussion. In comparing the performance of different criteria, \cite{ZhnEA23} state that it is not generally possible to foresee which criterion provides best out-of-sample predictive performance for a particular problem. The authors recommend performing a simulation study involving direct quantification of out-of-sample predictive performance using cross-validation, to quantify the relative performance of different information criteria on problems of the relevant type, as a rational basis for selecting the most appropriate criterion for that problem type. This suggestion was adopted by us in the current work to assess which of a range of criteria based on AIC, BIC, DIC and WAIC yields best performance in a specific extreme value prediction problem involving relatively small samples of {Coupled Model Intercomparison Project Phase 6 (CMIP6) output} for near-surface temperature \TA. 

{Desert regions (i.e. regions with high aridity and low precipitation) push the Earth's climate systems to its limits (see e.g. \cite{Wrn04}): observations of the Earth's most extreme temperatures occur in desert regions, but the response of extreme temperatures to climate forcing may also vary considerably between deserts. Studying predictions of the effects of climate change of extremes of desert temperatures is therefore a particularly pressing application. For example, \cite{ZhaEA21} report analysis of MODIS (Moderate Resolution Imaging Spectroradiometer) land surface temperature data for 2002-2019, examining spatio-temporal patterns of hot and cold extremes as well as DTR (diurnal temperature range). A maximum land surface temperature of 80.8$^\circ$C, observed in the Dash-e Lut Desert in Iran and the Sonoran Desert in Mexico, is $>10^\circ$C above the previous global record of 70.7$^\circ$C observed in 2005. The coldest temperature was observed in Antarctic at -110.9$^\circ$C. A maximum DTR of 81.8$^\circ$C is observed in a desert environment in China. Extreme temperatures in the world's desert regions are particularly interesting from a climate change perspective, as their arid land-atmosphere systems amplify warming responses to radiative forcing, increasing the signal-to-noise ratio of anthropogenic climate change and enabling earlier detection of climate signals relative to more temperate regions (\cite{Zho16}; \cite{HknEA20}). In addition, extreme desert heating influences large-scale atmospheric circulation through the strengthening of thermal lows, modulation of monsoon systems, and expansion of subtropical dry zones, with impacts extending well beyond arid regions (\cite{IPCC-SPM})}.

{Evaporative cooling in response to solar radiative forcing plays a central role in mitigating extreme Earth temperatures. The low moisture content of desert regions however provides the potential for greater surface and atmospheric heating. Nevertheless, the extreme temperature characteristics of specific desert regions depend on complex interactions of multiple physical processes on different spatial and temporal scales. For example, in the case of ``hot'' deserts, the Sahara is located in a subtropical high-pressure belt, where air in the descending branch of the Hadley Cell is compressed adiabatically and heated, creating a so-called ``heat low'' cap of warm, dry air preventing cloud formation. Death Valley in the Mojave Desert region is a narrow trough sitting below sea level. Air sinking into the valley from the surrounding mountains undergoes adiabatic compression and heating. Radiative ground heating causes rising air, trapped by valley walls, which then cools somewhat before sinking back into the valley to be further compressed and heated further. Australia's extreme heat is often driven by a stagnant continental high-pressure system that sits over the centre of the continent, cutting off any maritime air from the Southern Ocean. Moreover, the Simpson Desert exhibits a high concentration of parallel longitudinal red sand dunes supporting sparse spinifex vegetation. During droughts, vegetation dies, lowering albedo and allowing the red iron-rich sand to absorb more heat. Albedo is critical in the fundamentally different behaviour of a ``cold'' desert such as Antarctica. Ice and snow absorb only 10\%-20\% of solar radiation (in contrast to $>$ 50\% for sandy or rocky ``hot'' deserts), resulting in less surface heating and more reflected solar radiation. Near-surface air remains cold and dense, creating a permanent high-pressure cap that repels warmer air masses and prevents cloud formation. Further, sublimation of ice in cold deserts requires considerably more energy than evaporation of liquid water. Extreme temperatures in desert regions are also influenced by monsoons, but the exact interaction varies by desert region. For example, the West African Monsoon moderates daytime temperature highs in Saharan regions due to evaporative cooling, with increased humidity also moderating night-time lows. In the cold Antarctic desert, global atmospheric teleconnections for example link south-east Asian monsoons via Rossby waves with increased polar temperatures. In the Antarctic, the Polar Vortex and Antarctic Circumpolar Current (ACC) act as barriers to external heating, which are themselves influenced by global teleconnections such as ENSO (El Nino Southern Oscillation).}

{\cite{CokViz15} used reanalysis and observational data to estimate that the rate of surface temperature rise in the Sahara is two to four times than of the tropical mean over the period 1979-2012, driven by near-surface desert amplification of warming. Accompanying strengthening of the summertime heat low and African easterly jet, and weakening of the winter anticyclone is noted, as is the central role of the near-surface layer in both desert and polar amplification. \cite{GuoEA23} examines future trends in extreme high and low Sahara temperatures from CMIP6 output for four Shared Socio-Economic Pathways (SSP) scenarios. Similar growth in annual maxima and annual minima temperatures is reported under all except \SH (see Section~\ref{Sct:Dat}), for which it appears that annual maxima grow more quickly than annual minima. \cite{ZhoEA2024} presents evidence for a teleconnection in surface and tropospheric temperatures associated with the NAO over the Sahara and explore its spatio-temporal features.}

{\cite{WngEA25} use reanalysis data to interrogate recent Antarctic warming, estimating the relative impacts of thermodynamic and dynamic changes, and exploring the spatial variability of changes. \cite{TrnEA20} reports Antarctic temperature variability and change from station data, and note that phase of the Southern Annular Mode (SAM) exerts the greatest control of temperatures, together with tropical Pacific forcing and local atmospheric circulation variability at some locations. \cite{TrnEA21} analyse measured extreme Antarctic near-surface air temperatures, noting that the majority of record high temperatures were recorded after the passage of air masses over high orography, with the air being warmed by the foehn effect. At some stations in coastal East Antarctica the highest temperatures were recorded after air with a high potential temperature descended from the Antarctic plateau, resulting in an air mass more than 50$^\circ$ warmer than the maritime air. Record low temperatures at the Antarctic Peninsula stations were observed during winters with positive sea ice anomalies over the Bellingshausen and Weddell Seas. \cite{TngEA2026} report an unprecedented 2024 East Antarctic winter heatwave driven by Polar Vortex weakening and amplified by anthropogenic warming, impacting ice shelf stability and predictability of future Antarctic extremes. \cite{BrmEA26} uses observational reconstruction to examine Antarctic warming, revealing spatial structure in long-term temperature trends and disparities with CMIP6 model output.}

{In a non-stationary extreme value setting, generalised extreme value (GEV) regression is often used to relate variation of block maxima of some environmental variable with respect to covariates. In the current work, we apply parametric GEV regression to understand temporal changes in the distribution of annual maxima of \TA. In related work, \cite{KhrZwr05} uses parametric GEV regression, with linear variation of all GEV parameters in time, to quantify non-stationarity of annual maxima of various temperature indices. The regression is estimated using L-moments and maximum likelihood, and bootstrapping employed to provide a quantification of uncertainty. The authors show that spatially-averaged parameter estimates for GEV shape and scale show little change in time, whereas increases in temperature extremes can be predominantly explained by increases in the GEV location parameter: that is, global changes in temperature extremes tend to be associated primarily with the overall shift of distribution of annual extremes towards a warmer mean climate, and with only relatively small changes in the actual shape of the extreme temperature distribution. Changes in the distribution of extreme precipitation are more intricate. \cite{ZwrEA11} also uses GEV regression, now with non-stationary GEV location only, to quantify the influence on long return period daily temperature extremes at regional scales. \cite{KhrEA13} again uses GEV regression to update the analysis of \cite{KhrZwr05} for CMIP5 model outputs, estimated using both L-moments and maximum likelihood. Now, GEV location and scale are assumed to vary linearly in time, but GEV shape assumed time-invariant. \cite{KhrEA18} is a further follow-up paper, comparing CMIP5 and CMIP6 return values. \cite{LiEA21} examines annual extremes of daily temperature from CMIP6 models, by fitting stationary GEV models independently to each interval of a partition of the time domain into consecutive blocks. This is done, since the authors claim that it is not possible to fit non-stationary GEV models reliably to time series of climate model output with short length; we hope that the current work demonstrates that this claim is generally not true. \cite{AbdPpl23} use a mixture of model selection information criteria, including AIC, BIC and the Anderson-Darling statistic, to select models for extreme precipitation in CMIP6 projections. Models considered involve linear representations for GEV regression location and scale. No conclusions are made about the best choice of information criterion to use in their study; the output of all criteria are used for model selection, but details are not given.}

In this paper, we seek evidence for changing distributional characteristics of annual maxima and minima of \TA output {derived from daily data (see Section~2)} from five different CMIP6 GCMs, for some of the Earth's most prominent desert regions (characterised by low annual precipitation). We estimate non-stationary extreme value models in time, coupled over three forcing scenarios {(see Section~2)}, using Bayesian inference. Markov chain Monte Carlo (MCMC) for model fitting allows incorporation of prior information in estimation of the full joint (posterior) distribution of model parameters, and hence careful quantification of uncertainty. We consider candidate models encoding different extents of non-stationarity, and use carefully-validated model section criteria to choose optimal representations. We also demonstrate the potential for Bayesian model averaging (see e.g. \cite{Stl19}, \cite{Wss00}) as an alternative to model selection, following the ``stacking'' approach of \cite{YaoEA18}, using cross-validation to establish a weighting scheme of candidate models providing optimal predictive performance. 

Previous work by some of the current authors (e.g. \cite{EwnJnt23a} and \cite{LchEA25}) explored non-stationarity in the distributions of annual maxima and minima of GCM outputs of interest to the ocean engineering community. A series of articles including {the aforementioned} \cite{KhrZwr05}, \cite{ZwrEA11}, \cite{KhrEA13} and \cite{KhrEA18a} uses parametric GEV regression to quantify non-stationarity of annual maxima of various temperature indices. The parametric form of variation of GEV parameters in time for all of these articles are typically \emph{assumed}, rather than shown statistically to be optimal. Here we seek to \emph{justify} the functional form for variation of GEV parameters in time based on predictive performance. The impact of climate scenario on the change in the distribution of extreme temperatures has been examined by a number of authors, including \cite{KhrEA18b}. Research has tended to focus on independent analysis of data for different climate scenarios. However, for CMIP6 models we expect that the distribution of extreme \TA will be the same in the first year (2015) of output under all forcing scenarios. For this reason, here we consider coupled GEV regression models which ensure a common distribution of extreme \TA under all scenarios in 2015. A discussion of related work of interest, including software for GEV regression, and developments of semi-parametric extreme value regression, is provided in Section~SM1.2.

\subsection*{Objectives and outline}
Our objective is to develop a straightforward scheme for optimal model selection in extreme value regression, and use it to quantify a change in return value over some period, based on observations of annual maxima or minima of some process in time. {Without careful model selection, misleading conclusions regarding the characteristics of future extremes are likely. This article demonstrates and applies a simulation procedure to achieve reliable model selection}. We use the scheme to estimate the change, over the century (2025,2125) in the 100-year return value for annual maximum and minimum of \TA for the Earth's desert regions; see Section~\ref{Sct:Dat}. To estimate return values, we first characterise how the distribution of annual extrema varies in time using parametric GEV regression. Different candidate functional forms for the GEV model parameters (location, scale and shape; see Section~\ref{Sct:Mth:MdlCmp}) in time are considered. Model inference is performed using Bayesian inference. To select between candidate models, we use an information criterion. Since there are many information criteria in use for model selection (see Section~\ref{Sct:Mth:Slc}), and it is impossible before analysis to know which criterion is most appropriate to use, we perform a simulation study, generating samples with similar characteristics to those of the CMIP6 spatio-temporal extreme \TA output, to identify which information criteria provide best eventual estimates of the difference in return values; see Section~\ref{Sct:SmlMdlSlc}. Using the optimal information criteria, Section~\ref{Sct:Rsl} provides estimates of changes is return values. The discussion of Section~\ref{Sct:DscCnc} provides an initial assessment of estimates of differences in return values using Bayesian model averaging, exploiting a weighted sum of all candidate models instead of selecting a single candidate. 

In our application for a given GCM model ensemble, output for three forcing scenarios is available, corresponding to the same distribution of extreme temperatures in the first year of observation. We therefore estimate a \emph{coupled} GEV regression model imposing this constraint; see Section~\ref{Sct:Mth:GEVR}. Supplementary material (SM) provides supporting discussion and illustrations. Software and data for the analysis are provided at \cite{LchJnt26}.

\FloatBarrier
\section{Data}  \label{Sct:Dat}
%
Output for near-surface air temperature (\TA, {unit: }K) corresponding to five CMIP6 GCMs was accessed via the UK Centre for Environmental Analysis (CEDA) archive during the Spring of 2025. For each of these GCMs, gridded output for the whole globe is generally available daily for the time period 2015-2100. We examine output for three climate scenarios (or climate experiments): SSP126, SSP245 and SSP585. Each of these scenarios combines a Shared Socioeconomic Pathway (SSP) with a Representative Concentration Pathway (RCP). Experiment SSP126 follows SSP1, a storyline with low climate change mitigation and adaptation challenges, and RCP2.6 which leads to (additional) radiative forcing of 2.6 Wm$^{-2}$ by the year 2100. Analogously, experiment SSP245 (SSP585) follows SSP2 (SSP3), a storyline with intermediate (high) climate change mitigation and adaptation challenges, and RCP4.5 (RDC8.5) which leads to (additional) radiative forcing of 4.5 (8.5) Wm$^{-2}$ by the year 2100. 

\begin{figure}[!ht]
	\centering
	\includegraphics[width=1\textwidth]{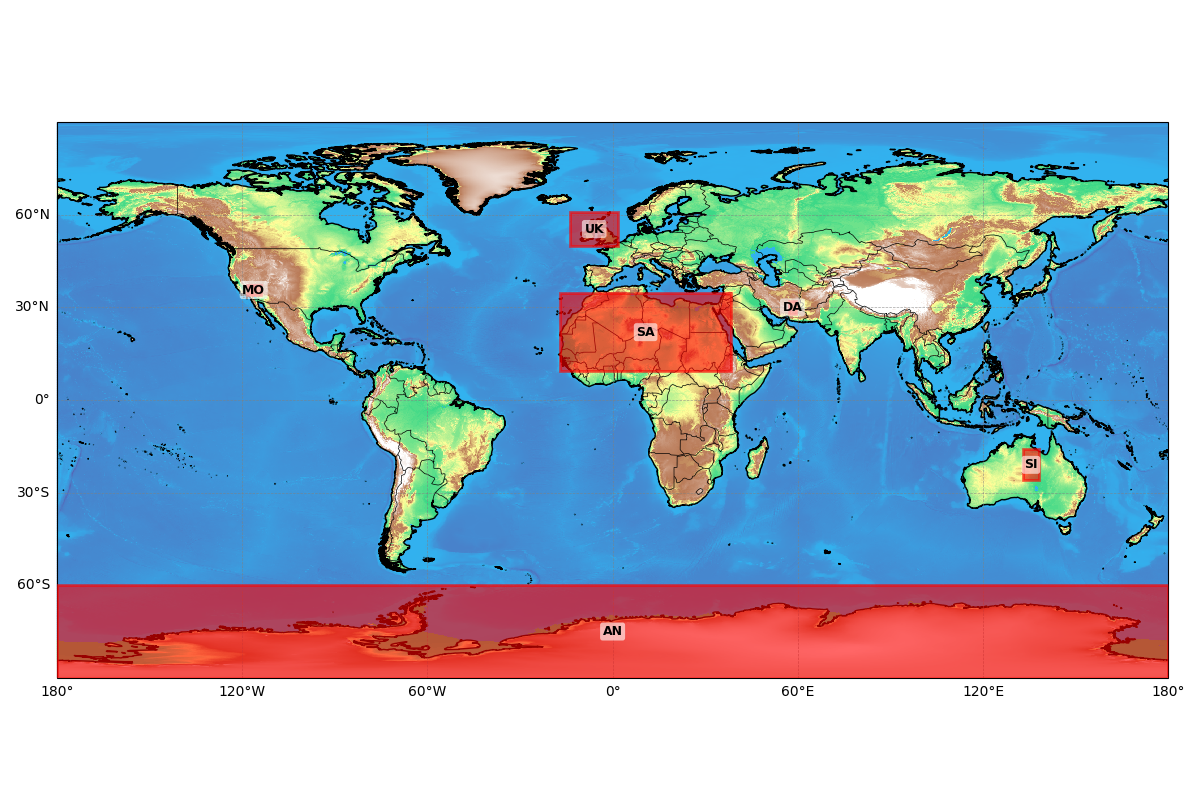}
	\caption{World map showing desert regions considered. Details of longitude-latitude bounding boxes in Table~\ref{Tbl:LctRgn}. Note that \DMO (MO) and \DDA (DA) are small geographic regions consisting of small numbers of GCM grid locations. {Hypsometric colour coding of bathymetry and orography is provided for general guidance only}.} 
	\label{Fgr-Map}
\end{figure}

Daily data are accessed directly from the CEDA archive, for the desert regions illustrated in Figure~\ref{Fgr-Map} and defined by the longitude-latitude bounding boxes of Table~\ref{Tbl:LctRgn}. {We also access corresponding data for the temperate UK region, which we henceforth refer to as a ``control'' region, in the sense that it provides a non-desert baseline, standard or comparator against which climate effects in desert regions can be judged.} For each region, we then calculated the annual maximum temperature, and the annual minimum temperature for the region as detailed in Section~\ref{Sct:Dat:SptTmpExt}.
Notice that the geographic areas of the different regions vary considerably, as does the number of GCM grid locations contained within each region. Regional maxima and minima for \DAN and \DSA in particular are taken over a large number of GCM locations in each case; inspection of Section~SM9 provides context. 

\begin{table}[!ht]
	
	\centering
		
		\begin{tabular}{|c|c|c|}
			\hline
			\multicolumn{1}{|c|}{\multirow{2}{*}{Region}} &
			\multicolumn{2}{|c|}{Interval \Tstrut} \\
			\cmidrule(lr){2-3}
			& Longitude & Latitude \Tstrut \\ 
			\hline
			
			\DAN (AN)\Tstrut & (-180$^\circ$, -180$^\circ$) & (-90$^\circ$,-60$^\circ$) \\ \hline
			\DDA (DA)\Tstrut & (57$^\circ$,59.5$^\circ$) & (28$^\circ$,32$^\circ$) \\ \hline
			\DMO (MO)\Tstrut & (-118.5$^\circ$,-114$^\circ$) & (34$^\circ$,37.25$^\circ$) \\ \hline
			\DSA (SA)\Tstrut & (-17.1$^\circ$,38.5$^\circ$) & (9.3$^\circ$,34.6$^\circ$) \\ \hline
			\DSI (SI)\Tstrut & (133$^\circ$,138$^\circ$) & (-26$^\circ$,-16$^\circ$) \\ \hline
			\DUK (UK)\Tstrut & (-13.7$^\circ$,1.8$^\circ$) & (49.9$^\circ$,60.8$^\circ$) \\ \hline

		\end{tabular}

	\caption{Summary of desert regions. List of geographic locations (and the UK as reference), with region acronyms in alphabetical order (column 1). Longitude-latitude bounding boxes per region (column 2).}
	\label{Tbl:LctRgn}
\end{table}

For each combination of desert region, GCM and climate scenario, we examine output for up to five climate model ensemble members (denoted r* in the climate nomenclature; see Table~\ref{Tbl:GcmDat} for examples) where available; these correspond to a common initialisation (i*), physics (p*) and forcing (f*) per GCM. For each combination of desert region, GCM, climate scenario and ensemble member, we therefore have access to time series, each consisting of 86 values, for annual maximum and annual minimum \TA for the period 2015-210. The choice of GCM to analyse was entirely pragmatic: we accessed output from all GCMs available on the CEDA archive at the time of data harvesting.

\begin{table}[!ht]
	
	\centering
	
	\begin{tabular}{|c|c|c|c|}
		
		\hline
		\multicolumn{1}{|c|}{\multirow{2}{*}{GCM}} &
		\multicolumn{1}{|c|}{\multirow{2}{*}{Ensemble reference}\Tstrut} &
		\multicolumn{2}{|c|}{{Resolution [$^\circ$]\Tstrut}} \\
		
		\cmidrule{3-4}
		
		& & {Longitude} & {Latitude}\Tstrut \\ \hline
		
		ACCESS-CM2 (AC)\Tstrut &
		r1i1p1f1, r2i1p1f1, r3i1p1f1, r4i1p1f1, r5i1p1f1 & {1.88} & {1.25}\\ \hline
		
		CESM2 (CE)\Tstrut &
		r4i1p1f1, r10i1p1f1, r11i1p1f1 & {1.25} & {0.94} \\ \hline
		
		EC-Earth3 (EC)\Tstrut &
		r1i1p1f1, r4i1p1f1, r11i1p1f1 & {0.70}  & {0.70} \\ \hline
		
		MRI-ESM2-0 (MR)\Tstrut &
		r1i1p1f1, r2i1p1f1, r3i1p1f1, r4i1p1f1, r5i1p1f1 & {1.13} & {1.13} \\ \hline
		
		UKESM1-0-LL (UK)\Tstrut &
		r1i1p1f2, r2i1p1f2, r3i1p1f2, r4i1p1f2, r8i1p1f2 & {1.88} & {1.25} \\ \hline
	\end{tabular}
	
	\caption{List of ensemble members considered per GCM. A total of five GCMs are used, listed in alphabetical order together with two-letter acronym (column 1). A total of up to five ensemble members is considered for each combination of climate variable and scenario (column 2), again depending on availability. Longitudinal and latitudinal GCM grid resolutions are also provided to two decimal places.}
	\label{Tbl:GcmDat}
\end{table}

\FloatBarrier
\subsection{Compilation of spatio-temporal summaries for desert regions}  \label{Sct:Dat:SptTmpExt}
\begin{figure}[!ht]
	\centering
	\includegraphics[width=0.49\textwidth]{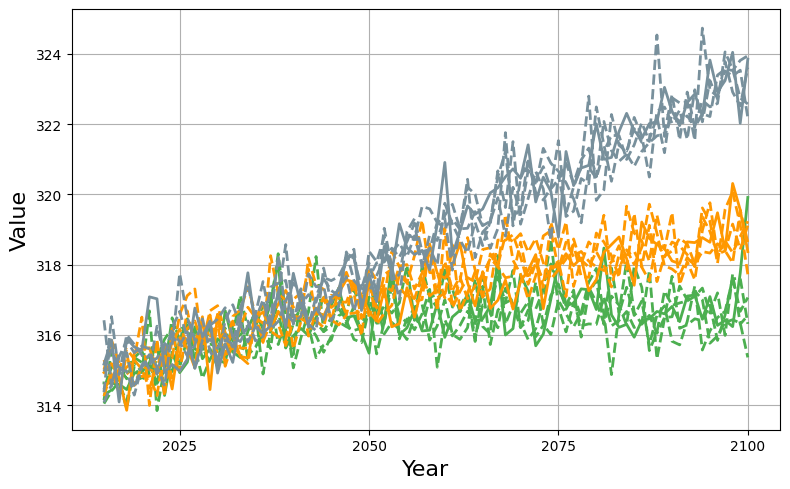}
	\includegraphics[width=0.49\textwidth]{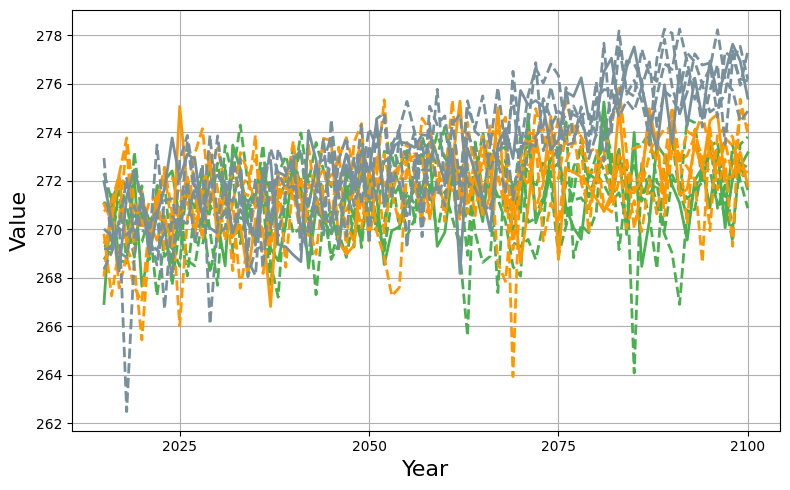}
	\caption{Regional annual maxima (left) and minima (right) time series of \TA (K) for the \DUK ``control'' region from the \UK GCM. Climate scenarios are distinguished by colour: \SL (green), \SM (orange), \SH (grey). Climate ensemble runs for each scenario, as listed in Table~\ref{Tbl:GcmDat}, are distinguished by line style.} 
	\label{Fgr-TS-SA-UK-MxmMnm}
\end{figure}

%
We consider analysis of spatio-temporal block maxima and minima of \TA, with temporal blocks corresponding to individual calendar years, and spatial blocks to each of the five desert regions (AN, DA, MO, SA and SI) and temperate UK control region. Henceforth, we refer to the spatio-temporal blocked data as ``regional annual maxima'' and ``regional annual minima'' for definiteness. For each combination of region, climate model, ensemble and scenario, regional annual maximum and minima data are extracted per calendar year from GCM output for the corresponding grid of locations (see Table~\ref{Tbl:LctRgn}). For illustration, Figure~\ref{Fgr-TS-SA-UK-MxmMnm} shows spatial annual maxima and minima time series for the \DUK region from \UK GCM. There is clear evidence for the effect of climate forcing, especially for maxima.

\begin{figure}[!ht]
	\centering
	\includegraphics[width=1\textwidth]{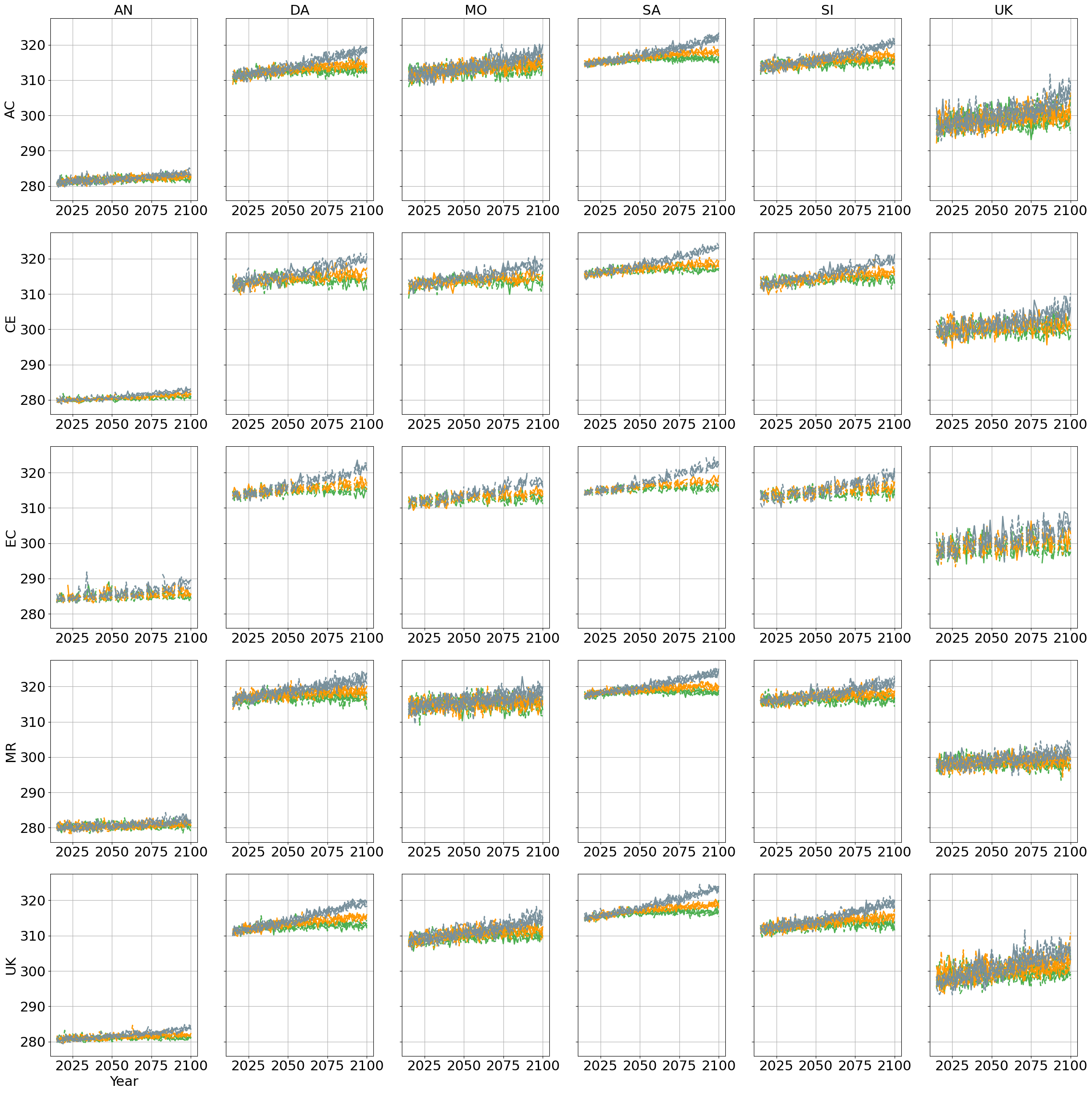}
	\caption{Time series of regional annual maxima of \TA (K) by GCM (rows: \AC, \CE, \EC, \MR and \UK) and region (column: : \DAN, \DDA, \DMO, \DSA, \DSI and \DUK). In each panel, climate scenarios are distinguished by colour: \SL (green), \SM (orange), \SL (grey). Climate ensemble runs for each scenario, as listed in Table~\ref{Tbl:GcmDat}, are distinguished by line style. Corresponding spatial annual minima time series are given in Figure~SM1. \ed{Note that the acronym ``UK'' refers to the \UK GCM in row headings, and to the United Kingdom location in column headings.}} 
	\label{Fgr-Mxm-TS-LctxGcm}
\end{figure}

Figure~\ref{Fgr-Mxm-TS-LctxGcm} shows regional annual maxima time series for all regions; corresponding regional annual minima time series are given in Figure~SM1. For both regional annual maxima and minima, differences between ``cold'', ``temperate'' and ``hot'' regions are clear, common to all GCMs. The extent of scenario effects varies somewhat by region and GCM for maxima in Figure~\ref{Fgr-Mxm-TS-LctxGcm}, but less so for minima in Figure~SM1. 

\FloatBarrier
\subsection{Calibration}  \label{Sct:Dat:Clb}
%
Outputs from GCMs are subject to bias. Therefore for interpretation, output of global climate models for variables such as \TA are typically transformed to regional scale projections using downscaling procedures, and potentially further calibrated using observations. Here we work directly with GCM output to quantify the extent to which return values for \TA change over the next century. We estimating the distribution of the difference between the 100-year return value for \TA in 2025 and 2125 directly. This approach avoids introducing extra variability in our estimates related to uncertain downscaling and calibration models, at the expense of potential bias due to lack of calibration. As noted in \cite{LchEA25}, successful calibration requires that reasonable data are available for both model output and direct measurement, over a relatively large proportion of the time period of interest, (2015,2125) here. We cannot be confident that a calibration developed for the years (2015,2025) (for which measurements are available) is appropriate in 2125. Note also that the \emph{difference} in 100-year return values for \TA over the next century is unaffected by constant bias correction calibration.

\section{Methodology}  \label{Sct:Mth}
%
We use parametric GEV regression to estimate models for regional annual maxima and minima of \TA, for regions defined in Section~\ref{Sct:Dat:SptTmpExt}. Generalising the approaches of \cite{EwnJnt23a} and \cite{LchEA25}, we assume that model parameters vary systematically over the period of observation as a function of time. Moreover, we constrain estimation across the three climate scenarios \SL, \SM and \SH such that the estimated tail in the first year of GCM output (2015) is common to all scenarios. We consider GEV regressions with different complexities, discussed in Section~\ref{Sct:Mth:MdlCmp}. Model estimation is outlined in Section~\ref{Sct:Mth:GEVR}. The methods we consider for model selection are discussed in Section~\ref{Sct:Mth:Slc}.

\subsection{Model complexities}  \label{Sct:Mth:MdlCmp}
%
Indexing climate scenarios \SL, \SM and \SH in terms of increasing forcing using variable $j$, $j=1,2,3$, we consider constant, linear and quadratic models for all of location $\mu$, scale $\sigma$ and shape $\xi$ of the GEV regression, of the form
\begin{eqnarray}
	\eta_j(t) = \eta_0  +  \tau \eta_{1j} + \tau^2 \eta_{2j}
	\label{Eqt:CvrTimQdr}
\end{eqnarray}
where $\eta \in \{\mu, \sigma, \xi\}$, for $i=1,2,...,P=86$, $j=1,2,3$, where $\tau = (i-1)/(P-1)$,  and calendar year $t=i+2014$. Constant and linear parameterisations are imposed by setting $\eta_{1j} = \eta_{2j} = 0$, and $\eta_{2j}=0$ respectively. For the location parameter $\mu$, we also consider an ``asymptotic'' parameterisation  
\begin{eqnarray}
	\mu_j(t) = \mu_0^a  + \mu^a_{1j} 
	\left(\frac{1 - \exp(-\tau/\mu^a_{2j})}{1 - \exp(-1/\mu^a_{2j})} \right) .
	\label{Eqt:CvrTimAsm}
\end{eqnarray}
The asymptotic parameterisation for $\mu$ is considered since intuitively it might be expected that, under some circumstances at least, climate forcing might cause a shift of \TA over time to a new stable level; it is reasonable to expect that this effect would be most identifiable for the location parameter. For the quadratic and asymptotic parameterisations, note that the values of $\eta_j(2015) = \eta_0$, $\eta \in \{\mu, \sigma, \xi\}$ and $\mu_j(2015)=\mu^a_0$, $j=1,2,3$ are common to all climate scenarios, imposing a common tail distribution for 2015. We also have $\eta_j(2100) = \sum_{k=1}^3 \eta_{kj}$ for the quadratic parameterisation, and $\mu_j(2100) = \sum_{k=1}^2 \mu^a_{kj}$. Note, for a given combination of GCM and region, different parametric forms are considered for each of $\mu$, $\sigma$ and $\xi$ as a function of time, but those forms are assumed common to all three climate scenarios.

We consider constant (C), linear (L), quadratic (Q) and asymptotic (A) parameterisations for GEV location $\mu$, and C, L and Q parameterisations for $\sigma$ and $\xi$. We refer to a specific model choice using an acronym indicating the parameterisations for $\mu$, $\sigma$ and $\xi$ in order. Thus, ``QLC'' indicates a GEV regression model with quadratic parameterisation for $\mu$, linear for $\sigma$ and constant for $\xi$. It is known that a sample is typically more informative for estimation of GEV location, then of scale then of shape (see e.g. discussion of \cite{JntEwnFrr08a}). For this reason, we can expect to identify more complex non-stationary behaviour in GEV location, then in scale, then in shape. As a result, we consider a total of 11 different GEV regression models, with the following complexities: CCC, LCC, QCC, ACC, LLC, LLL, QLC, QLL, QQC, QQL and QQQ. 

\FloatBarrier
\subsection{Coupled non-stationary generalised extreme value regression}  \label{Sct:Mth:GEVR}
%
Asymptotic extreme value theory (e.g. \cite{Cls01}, \cite{BrlEA04}, \cite{JntEwn13}) shows the distribution of block maxima, for random draws from a distribution within the maximum domain of attraction of a GEV distribution, can be approximated by a GEV for sufficiently large block size. With the three climate scenarios again indexed by variable $j$, $j=1,2,3$, we assume access to a sample $D=\{x_{ij}\}_{i=1,j=1}^{P,3}$ of regional annual maxima $X_j(t)$ ($i=1, 2, 3, ..., P;j=1,2,3; t=i+2014$) for analysis, for each combination of climate ensemble and region. Further we assume that for scenario $j$, the sample is drawn from a GEV distribution, with non-stationary location parameter $\mu_j(t) \in \mathbb{R}$, scale $\sigma_j(t)>0$, shape $\xi_j(t) \in \mathbb{R}$ and log density function
\begin{align}
	\log f_j(x) &= \log f_{\text{GEVR}}(x|\mu_j(t), \sigma_j(t), \xi_j(t)) \nonumber \\ 
	&=
	\begin{cases} 
		- \left[ 1 + (\xi_j(t)/\sigma_j(t))\left(x-\mu_j(t)\right)\right]^{-1/\xi_j(t)}  &\xi_j(t) \neq 0\\
		\exp(-(x-\mu_j(t))/\sigma_j(t)) &\xi_j(t) = 0 .
	\end{cases}
	\label{Eqt:GEVR:Dns}
\end{align}
Model parameters $\eta_j \in \{\mu_j, \sigma_j, \xi_j\}$ vary with calender year $t$ as described in Equations~\ref{Eqt:CvrTimQdr} and \ref{Eqt:CvrTimAsm}, such that a common extreme value tail is fitted for the first year $t=2015$. The $T$-year return value $Q_j(t)$ for scenario $j$ in calendar year $t$ is estimated as the $p=1-1/T$ quantile of the corresponding GEV distribution function $F_j(x)$, so that
\begin{align}
	Q_j(t) =
	\begin{cases} 
		(\sigma_j(t)/\xi_j(t)) \left[ \left(-\log p\right)^{-{\xi}_j(t)} -1\right] + {\mu}_j(t)  &\xi_j(t) \neq 0\\
		\mu_j(t)-\sigma_j(t) \log\left(-\log p\right) &\xi_j(t) = 0 .
	\end{cases}
	\label{Eqt:GEVR:RV}
\end{align}
Note that we only consider the case $T=100$ years in the current work for definiteness. Parameter estimation is performed using Bayesian inference as described in Section~SM3, with 
\begin{eqnarray}
	\ell(\un{\theta}) = \sum_{j=1}^3 \sum_{i=1}^{P} \log f_j(x_{ij}|\un{\theta})
	\label{Eqt:GEVR:SmpLkl}
\end{eqnarray} 
as sample log likelihood, for parameter vector $\un{\theta} = \{{\mu}_0,{\sigma}_0,{\xi}_0,\{{\mu}_{1j},{\mu}_{2j},
{\sigma}_{1j},{\sigma}_{2j},
{\xi}_{1j},{\xi}_{2j}\}_{j=1}^3\}$ (for the full quadratic model QQQ, and the analogous sets for the other 10 GEV regression models). MCMC inference (see Section~SM3) yields a sample $\{\un{\theta}_{(s)}\}_{s=1}^{n_S}$ of joint posterior estimates from $f(\un{\theta}|D)$ where $n_S=10,000$. These can be used to estimate the empirical distribution of quantities of interest, such as return values $Q_j(2025)$, $Q_j(2125)$, $j=1,2,3$, and the difference 
\begin{eqnarray}
	\Delta Q_j = Q_j(2125)-Q_j(2025)
	\label{Eqt:GEVR:DeltaQ}
\end{eqnarray}
in return value over the 100 years from 2025 to 2125. Note that extreme value analysis for regional annual minima is performed via characterisation of the right hand tail of the distribution of negated data.

\FloatBarrier
\subsection{Model selection}  \label{Sct:Mth:Slc}
%
We consider a large number of possible model parameterisations, and select between them using popular information criteria such as the Akaike information criterion (AIC), the Bayesian information criterion (BIC), the deviance information criterion (DIC) and the Wantanabe-Akaike Information Criterion (WAIC). 

AIC (\cite{Akk74}) is an approximation to the Kullback-Leibler divergence between the fitted and underlying true model forms, and takes the form $\text{AIC} = \mathcal{D}(\hat{\un{\theta}}) + 2 p$, where $p$ is the number of parameters estimated, and $\mathcal{D}(\un{\theta})$ is the sample deviance, defined by $\mathcal{D}(\un{\theta}) = -2 \ell(\un{\theta}) = -2 \log f(\un{y}|\un{\theta})$, evaluated at the maximum likelihood estimates $\hat{\un{\theta}}$ of $\un{\theta}$ maximising the sample log likelihood defined by Equation~\ref{Eqt:GEVR:SmpLkl}.  Despite its name, BIC (\cite{Sch78}) is typically used in a frequentist setting, but has been used in conjunction with Bayesian inference (e.g. \cite{NthEA12}). BIC is a large sample approximation to the Bayes factor, and defined by $\text{BIC} = \mathcal{D}(\hat{\un{\theta}}) + p \log n$ where $n$ is the number of observations for fitting. DIC (\cite{SpgEA02}, \cite{GlmEA13}) is typically recommended for model selection in Bayesian inference, and like AIC and BIC, takes the form of ``deviance estimate" plus ``complexity penality''. The \cite{SpgEA02} and \cite{GlmEA13} definitions of DIC are given in Table~\ref{Tbl:MdlSlc}, together with variants due to  \cite{vdL05} and \cite{And11}. \cite{GlmEA13} recommend use of WAIC (\cite{Wtn13}) to approximate out-of-sample predictive performance; WAIC again follows the same general form, as detailed in Table~\ref{Tbl:MdlSlc}: WAIC is thought to be a more fully Bayesian approach for estimating out-of-sample expectation, starting with the computed log pointwise posterior predictive density and then adding a correction for effective number of parameters to adjust for over-fitting.
\begin{table}[h!]
	\centering
	\resizebox{\columnwidth}{!}{%
		\begin{tabular}{|c|c|c|c|c|c|}
			\hline
			Abbreviation & Literature name & Lack of fit, A & Complexity penality, B & Reference\\
			\hline
			AIC1 & AIC & $\mathcal{D}(\hat{\un{\theta}})$ & $2 p$ & \cite{Akk74} \Tstrut\\
			AIC2 &  -   & $\mathcal{D}(\mathbb{E}_{\un{\theta}|D}[\un{\theta}])$ & $2p$ &  -\Tstrut\\
			AIC3 & Simplified BPIC & $\mathbb{E}_{\un{\theta}|D}[\mathcal{D}(\un{\theta})]$ & $2 p$ & \cite{And11} \Tstrut\\
			BIC1 & BIC & $\mathcal{D}(\hat{\un{\theta}})$ & $p \log n$ & \cite{Sch78} \Tstrut\\
			BIC2 &  -   & $\mathcal{D}(\mathbb{E}_{\un{\theta}|D}[\un{\theta}])$  & $p \log n$  &  -\Tstrut\\
			BIC3 &  -   & $\mathbb{E}_{\un{\theta}|D}[\mathcal{D}(\un{\theta})]$ & $p \log n$ &  -\Tstrut\\
			DIC1 & DIC & $\mathcal{D}(\mathbb{E}_{\un{\theta}|D}[\un{\theta}])$ & $2 p_{D1}$, Eqn~\ref{Eqt:pD1} & \cite{SpgEA02} \Tstrut\\
			DIC2 & DICalt & $\mathcal{D}(\mathbb{E}_{\un{\theta}|D}[\un{\theta}])$  & $2 p_{D2}$, Eqn~\ref{Eqt:pD2} &  \cite{GlmEA13}  \Tstrut\\
			DIC3 & Improved BPIC & $\mathbb{E}_{\un{\theta}|D}[\mathcal{D}(\un{\theta})]$ & $2 p_{D1}$, Eqn~\ref{Eqt:pD1} & \cite{vdL05}, \cite{And11} \Tstrut\\
			WAIC & WAIC & $-2$ lppd, Eqn~\ref{Eqt:lppd} & $2 \ \sum_{i=1}^n \text{var}_{\un{\theta}|D} \left[ \log f(y_i | \un{\theta}) \right]$ & \cite{Wtn13}  \Tstrut\\
			\hline
		\end{tabular}
	}
	\caption{Information criteria (of the form $A+B$) explored for model selection. Sample size is $n$, the number of model parameters $p$. $\mathcal{D}= -2 \log f(\un{y}|\un{\theta})$ is the sample deviance, or $-2$ times the sample log likelihood. $\hat{\un{\theta}}$ is the maximum likelihood estimate for $\un{\theta}$, and $\mathbb{E}_{\un{\theta}|D}[\un{\theta}]$ the posterior mean parameter estimate. $\mathbb{E}_{\un{\theta}|D}[\mathcal{D}(\un{\theta})]$ is the expected sample deviance under the posterior. Definitions of log pointwise predictive density lppd, and the complexity penalties $p_{D1}$ and $p_{D2}$ are given in the equations indicated, in the main text. BPIC refers to the Bayesian Predictive Information Criterion of \cite{And07}, intended as an improved version of DIC.}
	\label{Tbl:MdlSlc}
\end{table}
Expectations and variances under the posterior distribution are evaluated using the MCMC sample, such that for function $g(\un{\theta})$
\begin{eqnarray}
	\mathbb{E}_{\un{\theta}|D}[g(\un{\theta})] = \frac{1}{n_S} \sum_{s=1}^{n_S} g(\un{\theta}_{(s)}) \text{,   and   }
	\text{var}_{\un{\theta}|D}[g(\un{\theta})] = \frac{1}{n_S-1} \sum_{s=1}^{n_S} \left( g(\un{\theta}_{(s)}) - \mathbb{E}_{\un{\theta}|D}[g(\un{\theta})] \right)^2 .
	\label{Eqt:BIC2}
\end{eqnarray}
The log pointwise predictive density lppd is defined as
\begin{eqnarray}
	\text{lppd} = \sum_{i=1}^n \log \left( \frac{1}{n_S} \sum_{s=1}^{n_S} f(y_i | \un{\theta}_{(s)}) \right)  
	\label{Eqt:lppd}
\end{eqnarray}
and the penalties $p_{D1}$ and $p_{D2}$ for DIC are defined as
\begin{eqnarray}
	p_{D1} = \mathbb{E}_{\un{\theta}|D} [\mathcal{D}(\un{\theta})] - \mathcal{D}(\mathbb{E}_{\un{\theta}|D}[\un{\theta}]) 
	\label{Eqt:pD1}
\end{eqnarray}
and
\begin{eqnarray}
	p_{D2} = 2 \ \text{var}_{\un{\theta}|D} \left[ \log f(\un{y} | \un{\theta}) \right]  .
	\label{Eqt:pD2}
\end{eqnarray}

Inspired by the nature of the different combinations of ``deviance'' and ``complexity penalty'' terms associated with the different information criteria, we consider additional variants AIC2 and BIC2 of AIC and BIC utilising the posterior mean parameter estimate in place of the maximum likelihood estimate. In addition, variant BIC3 adapts BIC to use the posterior mean (predictive) deviance. Further, the analogous adaptation of AIC (to AIC3) results in the simplified BPIC introduced by Ando (and so could also be considered an extension of DIC). The original BPIC of \cite{And07} was developed to reduce over-fitting using DIC; a similar form of information criterion was previously proposed by \cite{vdL05}. The simplified BPIC of \cite{And11} is an approximation to BPIC, appropriate when the sample likelihood dominates the prior for well-specified candidate models. In comparing AIC, BIC and DIC, it might be suspected from Table~\ref{Tbl:MdlSlc} that the role of the complexity penalty is likely to be crucial, and that there is considerable similarity between some information criteria from nominally different classes.

Next, in Section~\ref{Sct:SmlMdlSlc}, we evaluate the performance of all information criteria in a simulation study using data with known characteristics mimicking those of the regional annual maximum and minimum \TA data. Subsequently, in Section~\ref{Sct:Rsl}, we adopt the best performing information criteria from the study, for model selection using the CMIP6 climate \TA data. 

\FloatBarrier\section{Simulation study to examine the predictive performance of different model selection criteria} \label{Sct:SmlMdlSlc}
%
As discussed in Section~\ref{Sct:Mth:Slc}, there is no reason to expect that different model selection criteria will choose the same models complexities (defined in Section~\ref{Sct:Mth}) for scenario-coupled GEV regression for a given application. For example, as can be seen from Table~\ref{Tbl:MdlSlc}, BIC penalises model complexity relatively severely, and might therefore be expected to produce more parsimonious models, possibly advantageous in the context of extreme value modelling.  

\begin{figure}[!ht]
	\centering
	\includegraphics[width=1\textwidth]{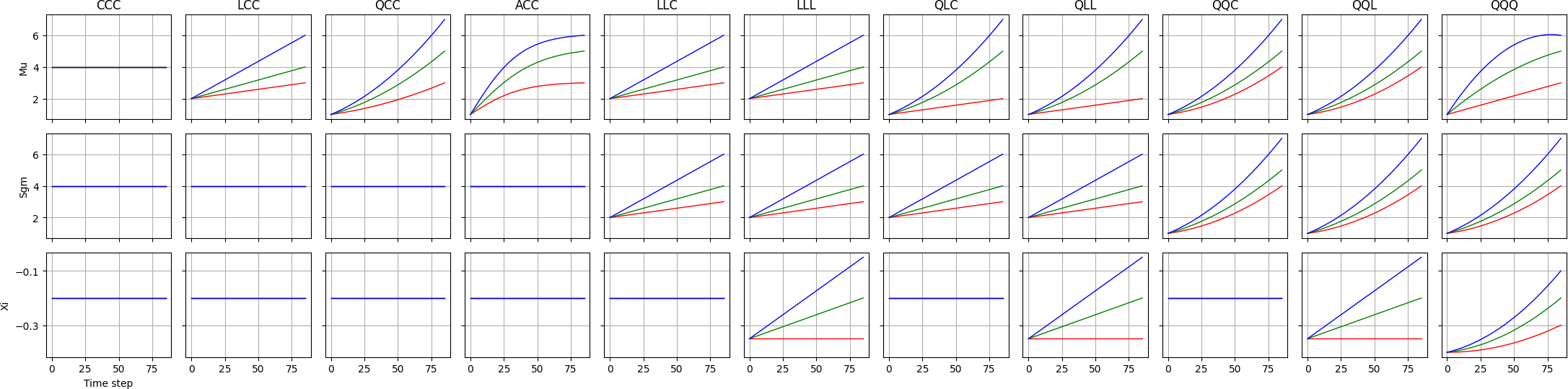}
	\caption{Simulation study design. Variation of GEV location ($\mu$; top), scale ($\sigma$; middle) and shape ($\xi$; bottom) in time for three scenarios (red, green and blue), for each of 11 data-generating processes (CCC to QQQ; columns left to right).} 
	\label{Fgr-Sml-Dsg}
\end{figure}

We perform a simulation study using the 11 data generating processes visualised in Figure~\ref{Fgr-Sml-Dsg}, one for each of the 11 model complexities listed in Section~\ref{Sct:Mth:MdlCmp}. The figure shows variation of GEV location, scale and shape parameters for three ``scenarios'' shown in red, green and blue; these choices of parameter variation were informed by inspection and initial analysis of the CMIP6 regional annual maximum and minimum data. We do not claim that the 11 examples used are representative of all possible non-stationary GEV processes, but they do reasonably reflect the characteristics of the CMIP6 data. We generate $n_R=100$ sample realisations $D_k$, $k=1,2,...,100$ from each of the 11 processes, each sample consisting of 86 ``yearly'' observations for each of three scenarios. We then fit all 11 model forms, ranging from CCC (i.e. constant GEV location, scale and shape) to QQQ (i.e. quadratic variation of GEV location, scale and shape in time) to each sample using MCMC, obtaining a chain of parameter estimates from the joint posterior distribution $f(\un{\theta}|D)$ (see Section~\ref{Sct:Mth:GEVR}). Then, for each combination of data-generating process and fitted model, we evaluate the root mean square error  
\begin{eqnarray}
	\text{RMSE} = (\frac{1}{3 n_R n_S} \sum_{j=1}^3 \sum_{k=1}^{n_R} \sum_{s=1}^{n_S} (\Delta Q_j^* - \Delta Q_{j(s)}|D_k])^2)^{1/2}
	\label{Eqt:RMSE}
\end{eqnarray}
in estimation of the difference $\Delta Q_j$ in return values over scenarios, $j=1,2,3$, where $\Delta Q_j^*$ is the true value, and $\Delta Q_{j(s)}|D_k$ is posterior estimate of $\Delta Q_j$ corresponding to iteration $s$, $s=1,2,...,n_S$ from the MCMC chain for data realisation $D_k$, $k=1,2,...,n_R$.

\begin{figure}[!ht]
	\centering
	\includegraphics[width=0.7\textwidth]{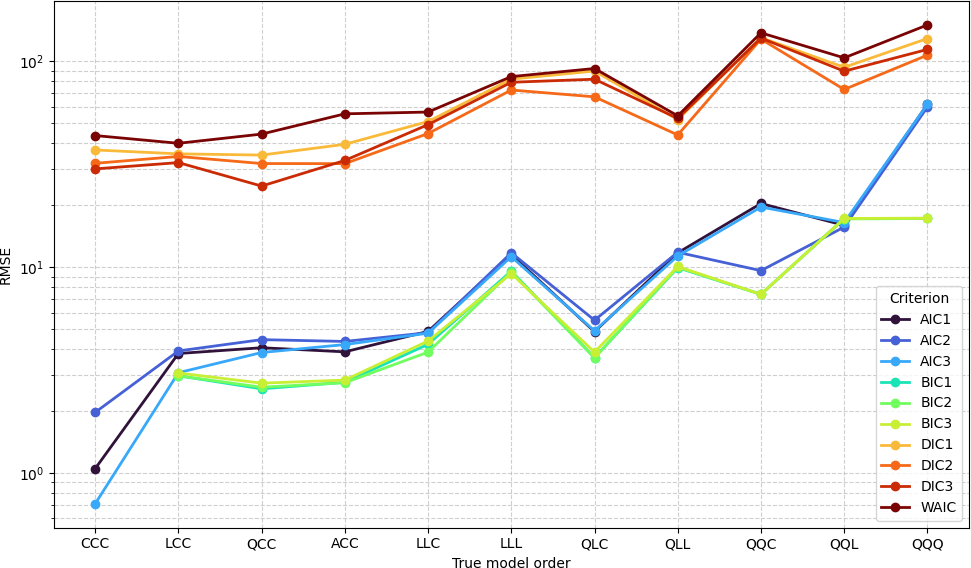}
	\caption{RMSE estimates (see Equation~\ref{Eqt:RMSE}) from the simulation study, for different combinations of data-generating process (CCC to QQQ; x-axis) and model selection information criterion (AIC1 to WAIC; coloured lines).  Note that for data-generating process CCC, all of BIC1, BIC2 and BIC3 always select the correct CCC model form, resulting in RMSE=0 (not shown on the $\log_{10}$-scale y-axis.)} 
	\label{Fgr-Sml-RMSE}
\end{figure}

Finally, we use each of the model selection information criteria in Table~\ref{Tbl:MdlSlc} to choose optimal model complexities, and corresponding estimates for RMSE; these are shown in Figure~\ref{Fgr-Sml-RMSE}, as a function of data-generating process and model selection information criterion. 
From the figure, it is obvious that variants of AIC and BIC perform considerably better than DIC variants and WAIC, and that BIC performs somewhat better than AIC. Moreover, there is little difference between the ``within-class'' performance of variants of AIC, of BIC, or of DIC. In hindsight, the poor performance of DIC and WAIC is probably to be expected (see e.g. the discussion of Section~7.3 of \cite{GlmEA14}, and Section~{SM1.1}). Given these results, we adopt one variant of BIC (BIC3) and one of AIC (AIC3) for model selection applied to the CMIP6 data in Section~\ref{Sct:Rsl}, noting that we could have adopted any of the BIC and AIC variants without loss of generality.

\FloatBarrier
\section{Results}  \label{Sct:Rsl}
%
In this section, we apply the GEV regression methodology from Section~\ref{Sct:Mth} and optimal model selection criteria BIC3 and AIC3 from Section~\ref{Sct:SmlMdlSlc} to estimate changes in 100-year return values over the period (2025,2125) for samples of regional annual maxima and minima of \TA, for different choices of desert region, GCM and climate ensemble. Results for regional annual maxima are summarised in Section~\ref{Sct:Rsl:Mxm} and Section~SM4, and for regional annual minima in Section~\ref{Sct:Rsl:Mnm} and Section~SM5. {Section 5.3 provides an assessment of the changing width of the distribution of extreme temperatures.}

\FloatBarrier
\subsection{Regional annual maxima}  \label{Sct:Rsl:Mxm}
\begin{figure}[!ht]
	\centering
	\includegraphics[width=1\textwidth]{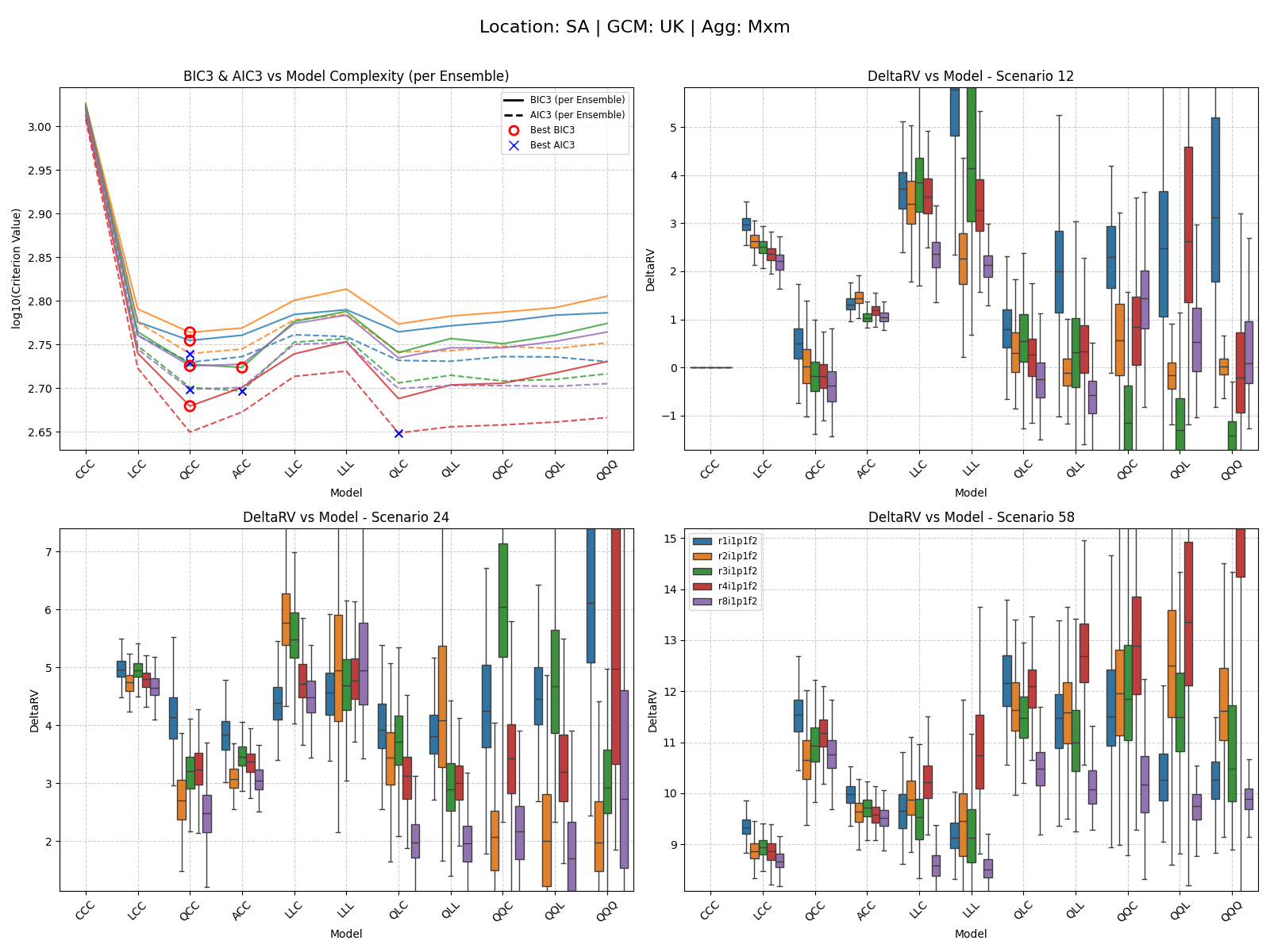}
	\caption{Summary of scenario-coupled GEV regression for regional annual maxima of the \DSA region using \UK GCM data. Top left: plots of BIC3 (solid line) and AIC3 (dashed line) for each available ensemble (distinguished by colour, see Table~\ref{Tbl:GcmDat} and legend in bottom-right panel); optimal model choice using BIC3 (AIC3) indicated using red disc (blue cross). Top right: box-whisker plots summarising the distribution of the difference in the 100-year return value between 2025 and 2125 ($\Delta Q_1$; see Equation~\ref{Eqt:GEVR:DeltaQ}) for climate scenario \SL as a function of fitted model complexity (x-axis) and ensemble (distinguished by colour, with consistent ensemble colouring across panels); location of horizontal centre line of each box indicates posterior median of $\Delta Q_1$; location of top (bottom) side of each box indicates 75\%ile (25\%ile) point, and top (bottom) of whiskers the 97.5\%ile (2.5\%ile) point of the posterior distribution. Bottom left and right: analogues of top right for scenarios \SM ($\Delta Q_2$) and \SH ($\Delta Q_3$). Value of $\Delta Q_j$, $j=1,2,3$ under model CCC is identically zero, and is omitted from bottom panels when convenient to provide better illustration of the variation in estimates under more complex models.} 
	\label{Fgr-Mxm-LctSA-GcmUK}
\end{figure}

%
A typical result for GEV regression for regional annual maxima, for a specific choice of region and GCM and all climate ensembles, is visualised in Figure~\ref{Fgr-Mxm-LctSA-GcmUK} (for the \DSA region using the \UK GCM). Results for all 30 combinations of region and GCM are summarised using the same template in Figures~SM2-SM31. The top left panel shows variation of values of information criteria (BIC3 and AIC3; different line styles) for fitting of models of different complexity (CCC to QQQ; x-axis), using data for each of the available climate ensembles (line colours). The optimal model selected is indicated by a red circle (BIC3) and blue cross (AIC3). The other three panels summarise the posterior distribution of the difference $\Delta Q_j$, $j=1,2,3$ in 100-year return value over the period (2025,2125) (see Equation~\ref{Eqt:GEVR:DeltaQ}) as a function of fitted model and climate ensemble, under climate scenarios \SL ($j=1$, top right), \SM ($j=2$, bottom left) and \SH ($j=3$, bottom right). 
For the \DSA region using the \UK GCM, the top left panel indicates that BIC3 prefers QCC models, with quadratic variation of GEV location in time, but stationary GEV scale and shape. AIC3 tends to prefer somewhat more complex models. The remaining panels show that the uncertainty in $\Delta Q$ increases with fitted model complexity, as does between-ensemble variability. For the BIC3-optimal QCC model fits, there is general agreement across climate ensembles: approximately no change in $\Delta Q$ under scenario \SL, but changes of around 3K and 11K under \SM and \SH. Clearly, the choice of fitted model has a large effect on our estimate of $\Delta Q$. The general features of results for other combinations of region and GCM (in Figures~SM2-SM31) are similar: BIC3 tends to select more parsimonious models, with LCC the most frequent choice (see Section~\ref{Sct:DscCnc:DstMdlCmp}). 

\begin{figure}[!ht]
	\centering
	\includegraphics[width=1\textwidth]{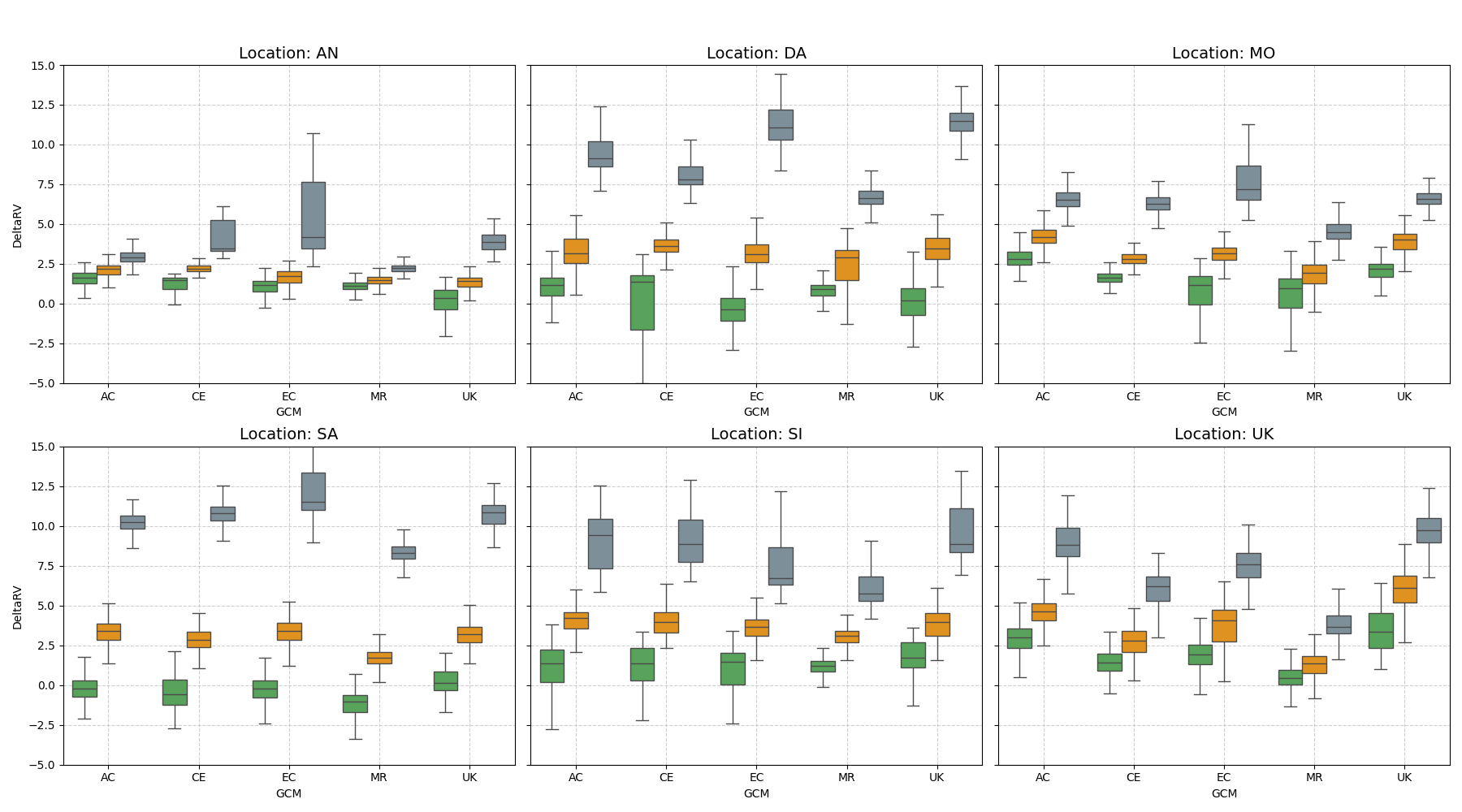}
	\caption{Summary of inferences for return value differences $\Delta Q_j$, $j=1,2,3$ (see Equation~\ref{Eqt:GEVR:DeltaQ}) of regional annual maxima using BIC3 for model selection, aggregated over climate ensemble. Each panel shows box-whisker plots summarising the posterior distribution of $\Delta Q$ for climate scenarios \SL ($j=1$, green), \SM ($j=2$, orange) and \SH ($j=3$, grey) and each of five GCMs (\AC, \CE, \EC, \MR, \UK). Left to right, top to bottom, panels show inferences for the \DAN, \DDA, \DMO, \DSA, \DSI and \DUK regions.} 
	\label{Fgr-Mxm-BW-LctGcmScn-BIC}
\end{figure}

\begin{figure}[!ht]
	\centering
	\includegraphics[width=0.7\textwidth]{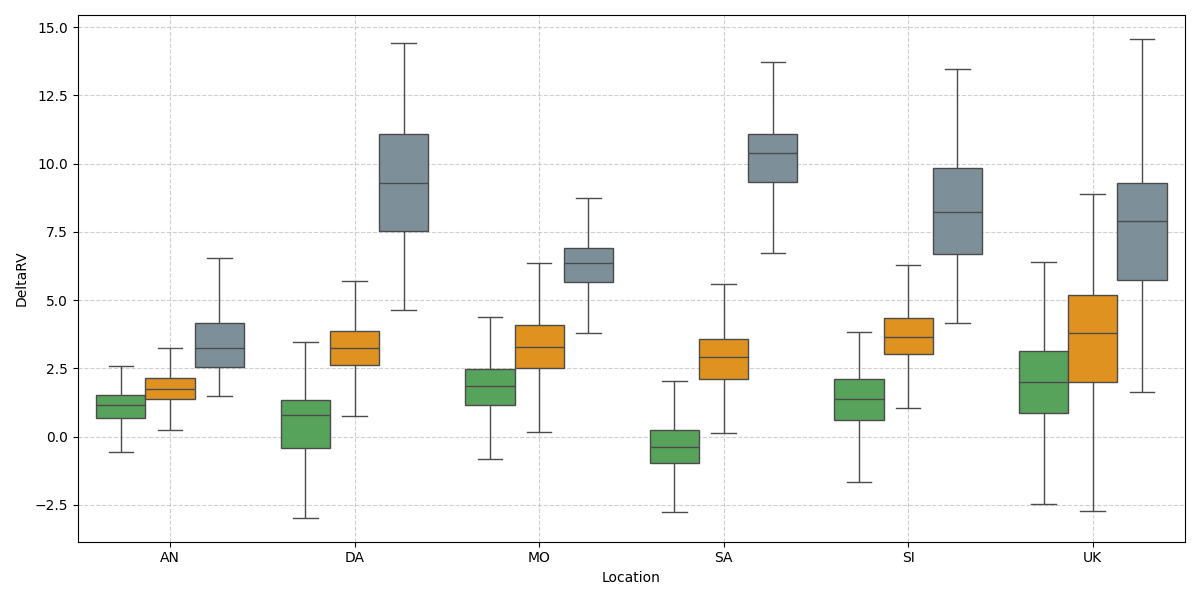}
	\caption{Summary of inferences for return value differences $\Delta Q_j$, $j=1,2,3$ (see Equation~7) of regional annual maxima using BIC3 for model selection, aggregated over climate ensemble and GCM. Box-whisker plots summarise the posterior distribution of $\Delta Q$ for climate scenarios \SL ($j=1$, green), \SM ($j=2$, orange) and \SH ($j=3$, grey) for the \DAN, \DDA, \DMO, \DSA, \DSI and \DUK regions.} 
	\label{Fgr-Mxm-BW-LctScn-BIC}
\end{figure}

Figure~\ref{Fgr-Mxm-BW-LctGcmScn-BIC} summarises inferences for $\Delta Q_j$, $j=1,2,3$ of regional annual maxima using the BIC3-optimal GEV regression models, aggregated over climate ensembles, as a function of region and GCM. Aggregation is achieve by simply combining the samples of posterior estimates for $\Delta Q$ corresponding to different ensembles into one aggregate sample. {In aggregating estimates for a particular quantity over climate models and ensembles, we assume that all estimates contributing to the aggregate are equally informative for that quantity}. The figure indicates general agreement across GCM and region, with somewhat larger increases in $\Delta Q$ in \DDA, \DSA and \DSI, and lowest in the \DAN. The \DUK control region does not look unusual alongside the ``hot'' desert regions. Corresponding estimates per climate ensemble are given in Figure~SM32, showing good general agreement across ensembles. The corresponding summaries for AIC3-optimal models is shown in Figures~SM33-SM34.
An overall summary for regional annual maxima is presented in Figure~\ref{Fgr-Mxm-BW-LctScn-BIC}, where the estimated posterior distribution of $\Delta Q_j$, $j=1,2,3$ using BIC3 for model selection is aggregated over both climate ensemble and GCM. The corresponding summary for AIC3 model selection is given in Figure~SM35.
It is noteworthy that the 95\% credible intervals for $\Delta Q$ under scenario \SL include zero (i.e. no change), whereas they do not under \SM and \SH, suggesting \emph{significant} increases in the 100-year return value for \TA over the next 100 years under the latter scenarios.

\FloatBarrier
\subsection{Regional annual minima}  \label{Sct:Rsl:Mnm}
%
Inferences for regional annual minima are presented using the same template used for maxima. Results for all 30 combinations of region and GCM are summarised Figures~SM36-SM65. The general characteristics of the figures are similar to those for regional annual maxima. The 100-year return value for \TA typically increases in time, but sometimes there is no obvious trend. The extent of any change observed is less marked. For example, for the \DSA and \UK GCM in Figure~SM55, BIC3 chooses LCC (rather than QCC), with typical changes of 2K, 4K and 7K under \SL, \SM and \SH. These differences reflect those visible in Figure~\ref{Fgr-TS-SA-UK-MxmMnm}.

\begin{figure}[!ht]
	\centering
	\includegraphics[width=1\textwidth]{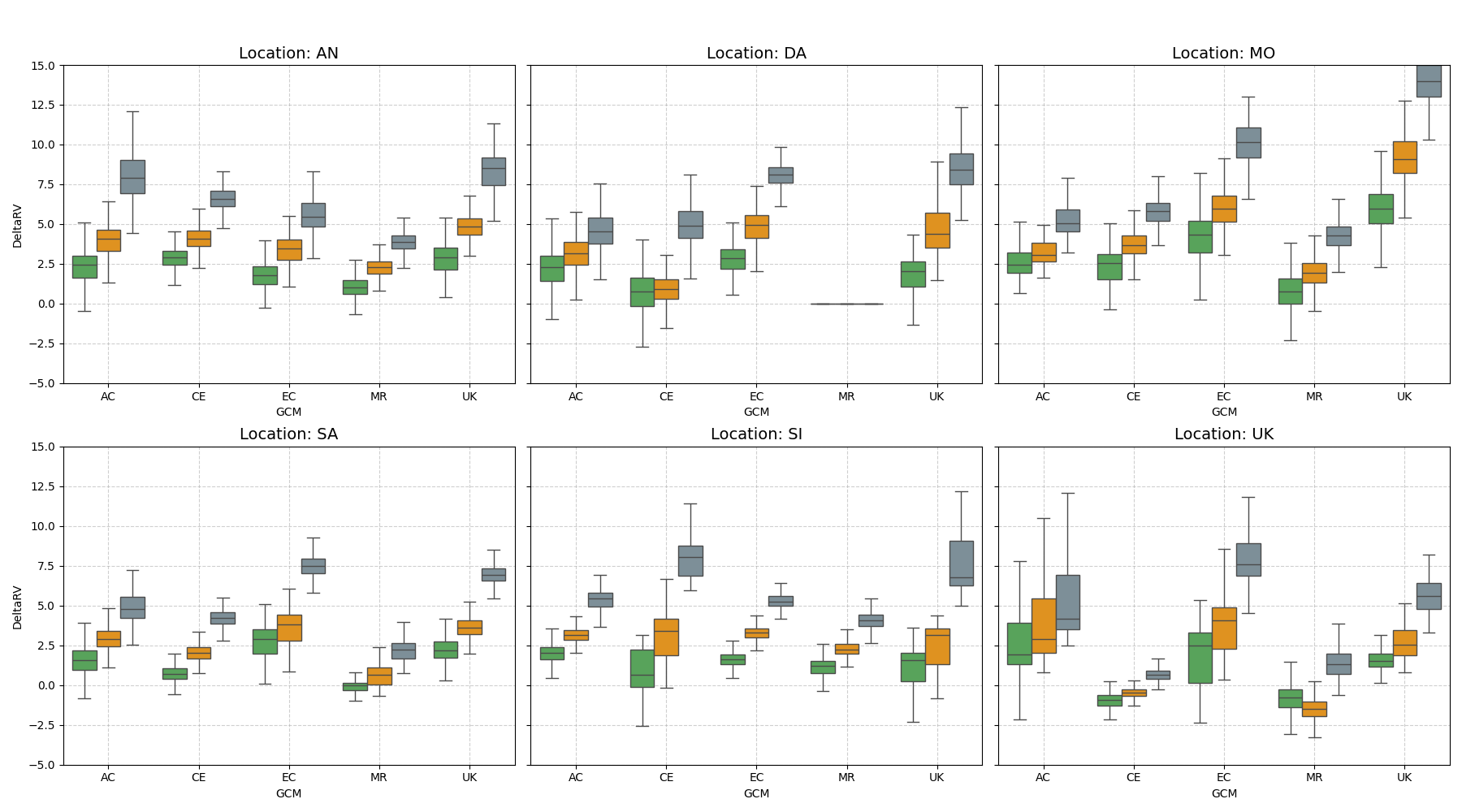}
	\caption{Summary of inferences for return value differences $\Delta Q_j$, $j=1,2,3$ (see Equation~\ref{Eqt:GEVR:DeltaQ}) of regional annual minima using BIC3 for model selection, aggregated over climate ensemble. For details, see Figure~\ref{Fgr-Mxm-BW-LctGcmScn-BIC}.}  
	\label{Fgr-Mnm-BW-LctGcmScn-BIC}
\end{figure}

\begin{figure}[!ht]
	\centering
	\includegraphics[width=0.7\textwidth]{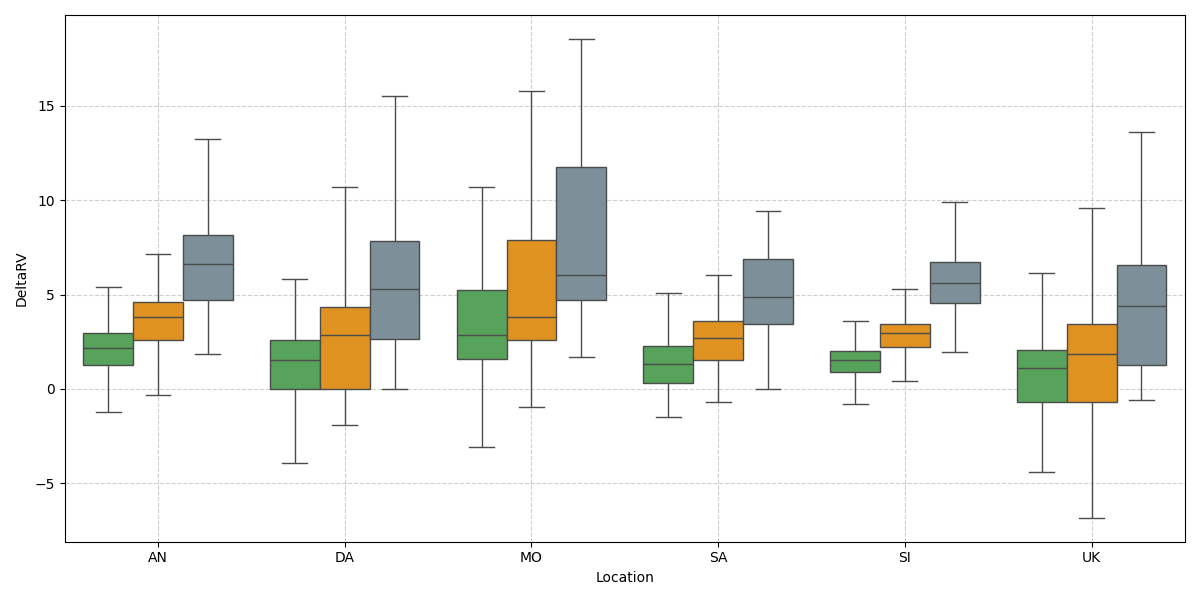}
	\caption{Summary of inferences for return value differences $\Delta Q_j$, $j=1,2,3$ (see Equation~7) of regional annual minima using BIC3 for model selection, aggregated over climate ensemble and GCM. For details, see Figure~\ref{Fgr-Mxm-BW-LctScn-BIC}.} 
	\label{Fgr-Mnm-BW-LctScn-BIC}
\end{figure}

Figure~\ref{Fgr-Mnm-BW-LctGcmScn-BIC} summarises inferences for $\Delta Q_j$, $j=1,2,3$ of regional annual minima using the BIC3-optimal GEV regression models, aggregated over climate ensembles, as a function of region and GCM; estimates per climate ensemble are given in Figure~SM66. The corresponding summaries for AIC3-optimal models is shown in Figures~SM67-SM68.
There is in general reasonable between-ensemble agreement (e.g. Figure~SM66), but poorer agreement between GCMs; for example, \MR yields CCC models for \DDA, resulting in $\Delta Q=0$ under all scenarios. Results for the \DAN are similar to those for ``hot'' regions. The overall summary for regional annual minima in Figure~\ref{Fgr-Mnm-BW-LctScn-BIC} gives estimates for the posterior distribution of $\Delta Q$ using BIC3 for model selection aggregated over both climate ensemble and GCM. It can be seen that many of the 95\% credible intervals include zero, even under \SM and \SH scenarios, indicating less confidence in the increasing trend observed. The corresponding summary for AIC3 model selection is given in Figure~SM69.

\subsection{Asymmetric warming: are extreme regional annual maxima increasing in size more quickly?} \label{Sct:Rsl:DltDltQ}
%
\begin{figure}[!ht]
	\centering
	\includegraphics[width=0.7\textwidth]{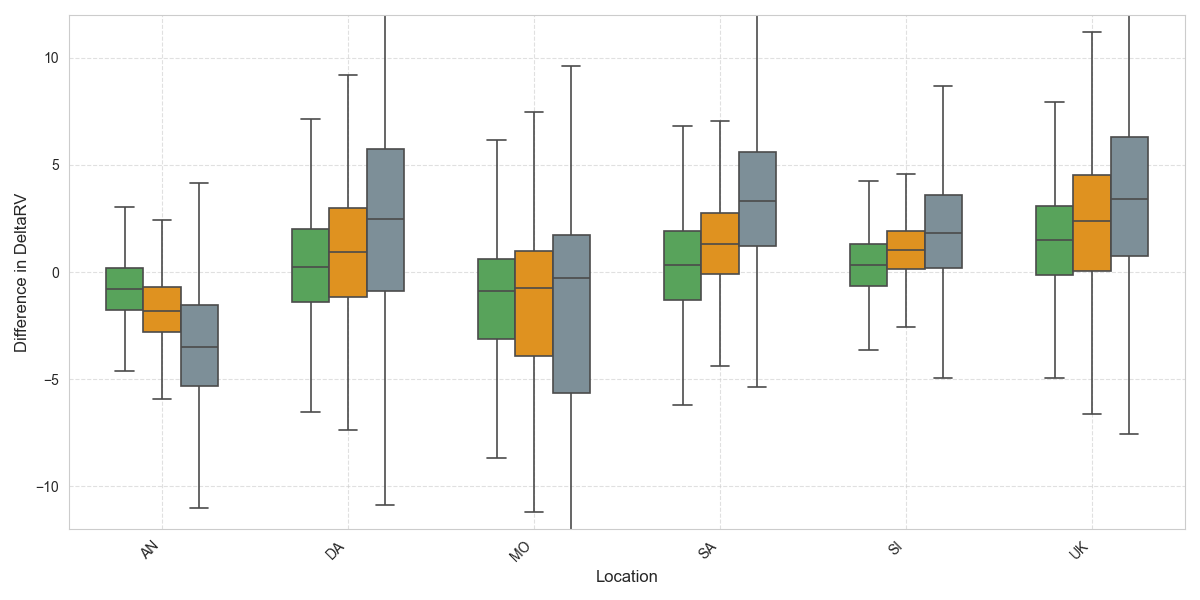}
	\caption{Summary of inferences for the difference $\Delta^2 Q_j$, $j=1,2,3$ (see Eqution~\ref{Eqt:DscCnc:DeltaDeltaQ}) between $\Delta Q$ values of regional annual maxima and minima. BIC3 was used for model selection, and posterior distributions aggregated over climate ensemble and GCM. For other details, see Figure~\ref{Fgr-Mxm-BW-LctScn-BIC}.} 
	\label{Fgr-MxmMnmDff-BW-LctScn-BIC}
\end{figure}

The distributions of $\Delta Q_j$, $j=1,2,3$ (see Equation~\ref{Eqt:GEVR:DeltaQ}) for regional annual maxima and minima are given in Figures~\ref{Fgr-Mxm-BW-LctScn-BIC} and \ref{Fgr-Mnm-BW-LctScn-BIC} respectively. It is apparent that there is in general an upward trend for both maxima and minima across desert and UK control regions, but that the extent of increase of regional annual maxima is in general greater. To quantify this, Figure~\ref{Fgr-MxmMnmDff-BW-LctScn-BIC} shows the differences $\Delta^2 Q_j$, $j=1,2,3$ for each combination of climate scenario and region, aggregated over climate ensemble and GCM. 
\begin{eqnarray}
	\Delta^2 Q_j = \Delta Q_j^\text{Mxm} -  \Delta Q_j^\text{Mnm}
	\label{Eqt:DscCnc:DeltaDeltaQ}
\end{eqnarray}
where $\Delta Q_j^\text{Mxm}$ and $\Delta Q_j^\text{Mnm}$ refer to $\Delta Q_j$ for regional annual maxima and minima respectively, $j=1,2,3$. {The 95\% credible intervals in Figure~\ref{Fgr-MxmMnmDff-BW-LctScn-BIC} include zero in every case, suggesting that there is no strong evidence for changes in $\Delta^2 Q$. Nevertheless, 75\% credible intervals (represented by the central coloured boxes per scenario) do not include zero for a number of regions under stronger forcing scenarios, providing some evidence for systematic change. For example in ``hot'' desert regions (and the temperate UK region), there is a mild increasing trend in median $\Delta^2 Q$ with increased forcing, suggesting that extreme quantiles of regional annual maxima are increasing (i.e. warming) at a greater rate than those of minima, and that this difference increases with climate forcing. Notably, the \DAN shows the opposite trend: under scenario \SH in particular, the value of median $\Delta^2 Q$ is around -3K: extremes of regional annual minima of \TA are increasing more quickly than those of regional annual maxima. The corresponding result for AIC3 model selection is given in Figure~SM70.}

{These features are quantified further in Table~\ref{Tbl:ExpPrb}, which gives empirical means for $\Delta Q$ (for regional annual maxima and minima) and $\Delta^2 Q$ for different combinations of region and climate scenario, aggregated over all climate models and ensembles. The table also provides empirical estimates of the probability that a given value (of $\Delta Q$ or $\Delta^2 Q$) is positive under the same aggregation. Hence, for example, under \SL, the empirical probabilities of $\Delta^2 Q>0$ are not extreme (i.e. near zero or unity) for any region. However, the probabilities become extreme under \SM and especially \SH, indicating significant asymmetry: a clear reduction in $\Delta^2 Q$ in the \DAN, but clear increase in the \DSA, \DSI and the temperate \DUK regions.} 
\begin{table}[!ht]
	\centering
	\color{lliw-ed} 
		\begin{tabular}{|c|c|c|c|c|r|r|r|}
			\hline
			\multicolumn{1}{|c|}{\multirow{2}{*}{Variable, $\Delta$\Tstrut}} &
			\multicolumn{1}{|c|}{\multirow{2}{*}{Region\Tstrut}} &
			\multicolumn{3}{|c|}{\(\mathbb{E}[\Delta]\)\Tstrut} &
			\multicolumn{3}{|c|}{\(\mathbb{P}(\Delta>0)\)\Tstrut} \\ \cmidrule(lr){3-8}
			& & SSP126 & SSP245 & SSP585 & SSP126 & SSP245 & SSP585\Tstrut \\ \hline
			\multirow{6}{*}{\begin{tabular}{c} $\Delta Q^{\text{Mxm}}$  \\ $[\text{K}]$ \end{tabular}} & AN\Tstrut & 1.01 & 1.72 & 3.55 & 0.90 & 1.00 & 1.00 \\    
			\cmidrule(lr){2-8}
			& DA\Tstrut & 0.41 & 3.16 & 9.32 & 0.68 & 0.99 & 1.00 \\
			& MO\Tstrut & 1.54 & 3.16 & 6.21 & 0.89 & 0.97 & 1.00 \\
			& SA\Tstrut & -0.38 & 2.85 & 10.24 & 0.34 & 1.00 & 1.00 \\
			& SI\Tstrut & 1.23 & 3.61 & 8.28 & 0.83 & 1.00 & 1.00 \\
			\cmidrule(lr){2-8}
			& UK\Tstrut & 2.05 & 3.69 & 7.70 & 0.93 & 0.99 & 1.00 \\
			\hline
			\multirow{6}{*}{\begin{tabular}{c} $\Delta Q^{\text{Mnm}}$  \\ $[\text{K}]$ \end{tabular}} & AN\Tstrut & 2.10 & 3.63 & 6.72 & 0.96 & 0.99 & 1.00 \\    
			\cmidrule(lr){2-8}
			& DA\Tstrut & 1.25 & 2.70 & 5.05 & 0.66 & 0.74 & 0.76 \\
			& MO\Tstrut & 3.42 & 5.10 & 8.05 & 0.94 & 0.99 & 1.00 \\
			& SA\Tstrut & 1.38 & 2.53 & 4.95 & 0.82 & 0.94 & 0.95 \\
			& SI\Tstrut & 1.38 & 2.81 & 5.89 & 0.90 & 0.99 & 1.00 \\
			\cmidrule(lr){2-8}
			& UK\Tstrut & 0.96 & 1.67 & 4.34 & 0.62 & 0.63 & 0.99 \\
			\hline
			\multirow{6}{*}{\begin{tabular}{c} $\Delta^2 Q$  \\ $[\text{K}]$ \end{tabular}} & AN\Tstrut & -0.71 & -1.75 & -3.69 & 0.29 & 0.09 & 0.03 \\    
			\cmidrule(lr){2-8}
			& DA\Tstrut & 0.41 & 0.73 & 2.08 & 0.54 & 0.61 & 0.73 \\
			& MO\Tstrut & -1.28 & -1.54 & -1.82 & 0.34 & 0.34 & 0.44 \\
			& SA\Tstrut & 0.30 & 1.30 & 3.35 & 0.55 & 0.80 & 0.91 \\
			& SI\Tstrut & 0.30 & 1.06 & 1.89 & 0.62 & 0.86 & 0.93 \\
			\cmidrule(lr){2-8}
			& UK\Tstrut & 1.49 & 2.28 & 3.47 & 0.85 & 0.88 & 0.88 \\
			\hline
		\end{tabular}%
	
	\caption{{Estimated expected \ed{value ($\mathbb{E}$) of} change and probability \ed{($\mathbb{P}$)} of increase (over the period (2025,2125)) for $\Delta Q^{\text{Mxm}}$ (for annual regional maxima), $\Delta Q^{\text{Mnm}}$ (for annual regional minima) and $\Delta^2 Q$ per region and climate scenario, estimated using output of optimal GEV models aggregated over climate model and ensemble.}}

\label{Tbl:ExpPrb}

\end{table}

{There are many competing potential physical drivers of the changes in $\Delta Q^{\text{Mxm}}$, $\Delta Q^{\text{Mnm}}$ and $\Delta^2 Q$ summarised in Table~\ref{Tbl:ExpPrb}. We are dealing with changes in regional annual extrema, likely to be affected by changes in temporal (e.g. seasonal) or spatial (e.g. coastal, altitudinal) effects. For our hot and cold desert regions, variation in evaporative cooling, cloud cover, soil moisture - albedo feedback, aridification, adiabatic sinking, atmospheric dust, aerosols and greenhouse gas concentrations, atmospheric column depth, and the changing influence of large-scale external drivers could all contribute to differences in near-surface temperature response to climate forcing, and asymmetric warming trends. In a hot desert, we speculate that asymmetric warming (with $\Delta^2Q>0$) with increasing climate forcing might be explained by the dominant role of radiative forcing over evaporative cooling on daytime heat, and low thermal inertia to maintain associated nighttime warming. In a cold desert, asymmetric warming (with $\Delta^2Q<0$) with increasing climate forcing might indicate breakdown of surface inversion in winter, with atmospheric turbulence mixing warmer air from above. Latent heat of fusion also provides a thermal buffer to increases of maximum Antarctic temperatures above 0$^\circ$C under climate forcing, not impact increases in minimum temperatures. As discussed in Section~\ref{Sct:Int}, for both hot and cold deserts, large-scale influences (e.g such as expansion of the Hadley Cell for the Sahara, or variations of the Polar Vortex for Antarctica) are possible contributors. That the drivers of these changes are complex is likely, given studies such as that of \cite{MatEA16}, which find complex spatio-temporal asymmetric trends in seasonal temperature variability in European instrumental records from 1864 to 2012, for both temperature maxima and minima. \cite{ZhnEA23a, LiuEA24} also report complex trends in diurnal temperature range, and reversed asymmetric warming of sub-diurnal temperature over land.}

\FloatBarrier
\section{Discussion and conclusions}  \label{Sct:DscCnc}

This paper presents is a pragmatic approach to model selection in Bayesian extreme value regression. We use the approach to estimate the change in the value of the 100-year return value $\Delta Q$ for near-surface atmospheric temperature \TA over the period (2025,2125), and its variability from CMIP6 model output for desert regions. Model fitting is performed using non-stationary generalised extreme value (GEV) regression and Bayesian inference. The GEV parameters $\mu, \sigma, \xi$ are assumed to exhibit parametric variation with time. Models for $\mu, \sigma, \xi$ in time with different complexities are considered and their performance quantified and compared using variants of the Bayesian and Akaike information criteria. These specific choices of criteria were found to be optimal in terms of minimising the root mean square error of predictions of $\Delta Q$ in a simulation study. We believe the work provides a number of interesting statistical and physical insights.

\subsection{Distribution of complexity of fitted models} \label{Sct:DscCnc:DstMdlCmp}
\begin{figure}[!ht]
	\centering
	\begin{subfigure}[b]{0.5\textwidth}
		\centering
		\caption*{Regional annual maxima}
		\includegraphics[width=1\textwidth]{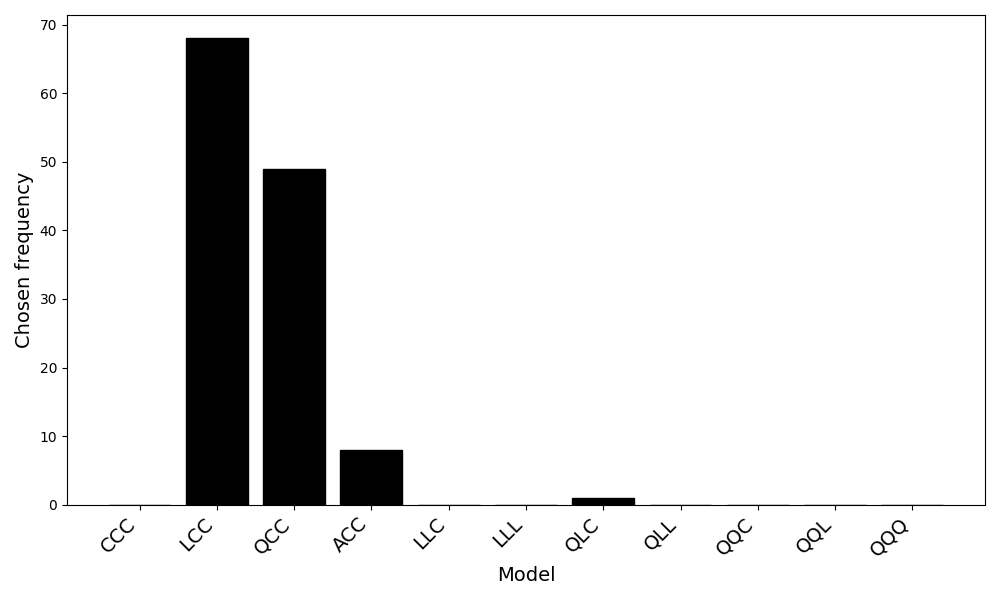}
	\end{subfigure}%
	\begin{subfigure}[b]{0.5\textwidth}
		\centering
		\caption*{Regional annual minima}
		\includegraphics[width=1\textwidth]{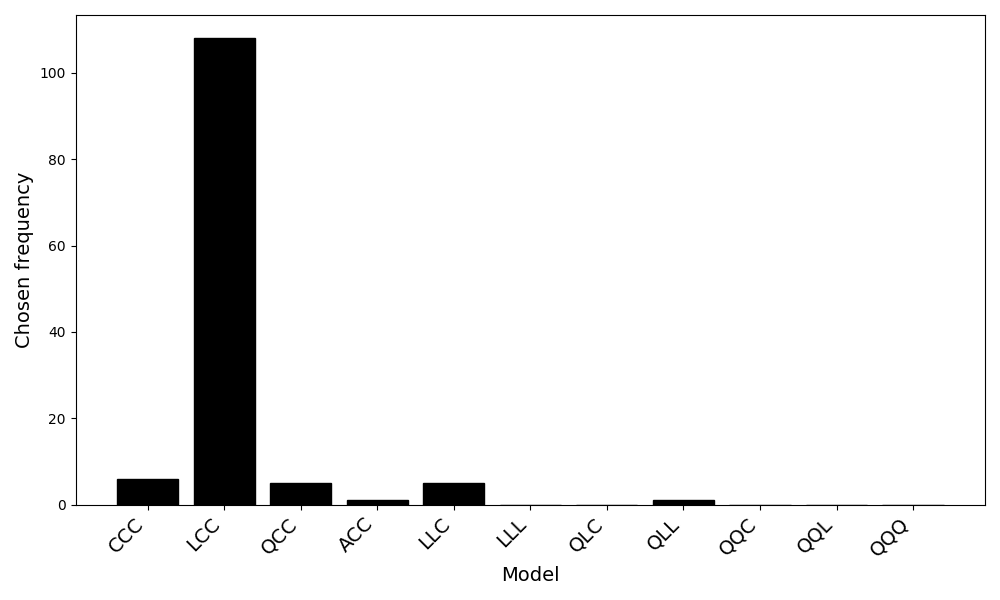}
		
	\end{subfigure}%
	\caption{Complexities of optimal fitted models for regional annual maxima (left) and minima (right) using BIC3 for model selection.} 
	\label{Fgr-Hst-MdlOrd-BIC}
\end{figure}

%
Numerous model selections are performed in the current analysis. The complexity of optimal model selections over all combinations of region, maximum or minimum, GCM and climate ensemble provides a general indication of the information content of the time series data. For example, the histograms in Figure~\ref{Fgr-Hst-MdlOrd-BIC} indicate the frequency with which each of the 11 candidate model complexities is chosen, in application to regional annual maxima (left hand side) and minima, over all combinations of region, GCM and climate ensemble, using BIC3 for model selection.
{Comparison of the left and right hand sides of the figure show that more complex models can be justified for regional annual maxima. GEV models with linear, quadratic and ``asymptotic'' functional forms in variation of location parameter with time, provide optimal predictions of change in 100-year return value for regional annual maxima. However, in general, selected models tend to incorporate non-stationarity in the GEV location parameter $\mu$ only, and very rarely in GEV scale $\sigma$ and shape $\xi$. For regional annual maxima, models QCC and ACC together account for almost half the selected model forms. For comparison, Figure~SM71 shows the corresponding histograms for model selection using AIC3. Relative to BIC3, model selection using AIC3 results in more complex model forms; regardless, models stationary with respect to $\sigma$ and $\xi$ are more prevalent. For regional annual minima with BIC3, there are cases demanding linear trends in both GEV location and scale parameters. Allowing variation of the GEV shape parameter is never justified for the application data examined and RMSE scoring rule adopted in this work}.

\subsection{An initial comparison with Bayesian model averaging} \label{Sct:DscCnc:BMA}
%
The current work has focussed on model selection from a set of 11 candidate models CCC to QQQ with different complexities. An alternative approach, avoiding selection of a single optimal model complexity for prediction of $\Delta Q$, it to estimate a weighted sum of all candidates which has good predictive performance for $\Delta Q$. Here, we provide an initial comparison of our model selection procedure with Bayesian model averaging (BMA).

In BMA, we take a linear combination of posterior predictive densities for $\Delta Q$ from each of $n_M=11$ candidate models $\{$CCC, LCC, ..., QQQ$\}$ $=\{M_1, M_2, ..., M_{11}\}$ in order for convenience. Specifically, for the 3-vector $q=[\Delta Q_1, \Delta Q_2, \Delta Q_3]$, we estimate the average predictive posterior density for $q$ as  
\begin{eqnarray} \label{Eqt:SQGY}
	f_S(q|D)=\sum_{m=1}^{n_M} w_m f(q|M_m, D)
\end{eqnarray} 
where $f(q|M_m, D)$ is the posterior predictive density for $q$ estimated from data $D$ using model $M_m$, $m=1,2,...,n_M$. We estimate the set of weights $\{w_m\}$ using the stacking procedure of \cite{YaoEA18} outlined in Section~SM8.

We compare the performance of Bayesian model averaging to model selection using BIC3, in application to the estimation of $\Delta Q_j$, $j=1,2,3$ using the simulation study data from Section~\ref{Sct:SmlMdlSlc}, and the CMIP6 climate data for the \DSA region and \UK GCM.
\begin{table}[]
	\resizebox{\columnwidth}{!}{%
		\begin{tabular}{|rr|r|r|r|r|r|r|r|r|r|r|r||r||r|}
			\hline
			&   & \multicolumn{12}{|c||}{\textbf{Data-generating model ($n_R=100$ realisations)\Tstrut}} & \multicolumn{1}{|c|}{\textbf{CMIP6 data\Tstrut}}\\
			\cmidrule{3-15}
			&	& CCC\Tstrut & LCC  & QCC  & ACC & LLC  & LLL  & QLC  & QLL  & QQC  & QQL  & QQQ & \multicolumn{1}{|c||}{LCC} & \multicolumn{1}{|c|}{SA-UK}   \\
			\cmidrule{3-15}
			&   & \multicolumn{11}{|c||}{Proportions of fitted models selected using BIC3\Tstrut} & \multicolumn{2}{|c|}{BMA weights $w_m$, $m=1,2,...,11$} \\
			\cmidrule{3-15}
			\hline
			\multirow{11}{*}{\rotatebox[origin=c]{90}{\textbf{Fitted model}}} 
			&	\multicolumn{1}{|c|}{CCC\Tstrut} & 1.00 & 0.68 & 0.10 & 0.18 & 0.05 &      &      &      &      &      &      & 0.14 (0.00,0.59) & 0.00  \\
			&	\multicolumn{1}{|c|}{LCC}        &      & 0.31 & 0.90 & 0.78 & 0.01 &      & 0.02 &      &      &      &      & 0.40 (0.00,1.00) & 0.00  \\
			&	\multicolumn{1}{|c|}{QCC}        &      &      &      & 0.03 &      &      &      &      &      &      &      & 0.00 (0.00,0.79) & 0.21  \\	
			&	\multicolumn{1}{|c|}{ACC}        &      &      &      & 0.02    &      &      &      &      &      &      &      & 0.18 (0.00,0.91) & 0.42 \\
			&	\multicolumn{1}{|c|}{LLC}        &      & 0.01 &      &     & 0.94 & 0.93 & 0.98 & 0.97 & 0.99 & 0.84 & 0.94 & 0.07 (0.00,0.74) & 0.00  \\	
			&	\multicolumn{1}{|c|}{LLL}        &      &      &      &     &      & 0.07 &      & 0.03 &      & 0.13 & 0.06 & 0.04 (0.00,0.46) & 0.00 \\
			&	\multicolumn{1}{|c|}{QLC}        &      &      &      &     &      &      &      &      & 0.01 & 0.03 &      & 0.01 (0.00,0.21) & 0.25 \\
			&	\multicolumn{1}{|c|}{QLL}        &      &      &      &     &      &      &      &      &      &      &      & 0.01 (0.00,0.00) & 0.09 \\
			&	\multicolumn{1}{|c|}{QQC}        &      &      &      &     &      &      &      &      &      &      &      & 0.01 (0.00,0.12) & 0.00 \\
			&	\multicolumn{1}{|c|}{QQL}        &      &      &      &     &      &      &      &      &      &      &      & 0.01 (0.00,0.12) & 0.00 \\
			&	\multicolumn{1}{|c|}{QQQ}        &      &      &      &     &      &      &      &      &      &      &      & 0.03 (0.00,0.36) & 0.02 \\
			\hline
			& RMSE\Tstrut                        & 0.00 & 3.06 & 2.82 & 2.73 & 4.38 & 9.30 & 3.86 & 10.1 & 7.38 & 17.3 & 17.3 & 3.98 &  - \\ 
			\hline
		\end{tabular}
	}
	\caption{{Assessment of fitted models for three analyses: (a) single model selection using BIC3 from simulated data (results in columns 3-13); (b) BMA from simulated data (results in column 14); and (c) BMA for CMIP6 sample of regional annual maxima for the \DSA from the \UK GCM. For analysis (a), columns 3-13 give the proportions of selected models (from the 11 candidates CCC to QQQ; table rows) using the BIC3 information criterion, corresponding to each of $n_R=100$ sample realisations from data-generating models with different complexities (CCC to QQQ; table columns). For analysis (b), column 14 gives the corresponding average BMA weights per fitted model, estimated from the same $n_R=100$ sample realisations of the LCC model only, with marginal 95\% uncertainty bands. For analysis (c), column 15 gives BMA weights, averaged over climate ensemble, for the CMIP6 sample data . Values are rounded to two decimal places, and zero values of proportions are not shown to aid interpretation. (Note the curious BMA weight estimate for model QLL, arising because values for all but one realisation are zero to two decimal places.)}}
	\label{Tbl:BMA}
\end{table}
Table~\ref{Tbl:BMA} summarises the distribution of selected model complexities in BIC3-based model selection for different data-generating models. Specifically, the same $n_R=100$ realisations of each of the $11$ data-generating models CCC to QQQ, reported in the simulation study of Section~\ref{Sct:SmlMdlSlc}, were re-used. Each of the $n_M$ candidate models CCC to QQQ is fitted in turn, and BIC3 used to select the optimal fitted model. Columns 3-13 of the table report the proportion of fitted models of different complexities, for each complexity of data-generating model: thus, for an LLL data-generating model, the LLC fitted model was selected by BIC3 for 93 of the 100 sample realisations, and LLL for the remaining 7 realisations. It can be seen that realisations of stationary samples (CCC) always result in selection of stationary (CCC) fitted models. When sample realisations include non-stationary variation (i.e. either linear, L or quadratic, Q) of GEV location $\mu$ or scale $\sigma$, this typically results in the selection of a fitted model with \emph{linear} variation (L) in the corresponding parameter. Fitted models incorporating variation in GEV shape $\xi$ are rarely selected, regardless of the data-generating process. Hence, the optimal fitted model complexity corresponding to all the data-generating models LLL to QQQ is LLC. The fact that fitted models with quadratic (Q) variation in $\mu$ and $\sigma$, and any non-stationary (i.e. L or Q) with respect to $\xi$, tend \emph{not} to be fitted, is attributed to the small sample sizes (86 years $\times$ three scenarios) in the current work. It is interesting that the typical literature presumption for parametric GEV regression applied to annual maxima of \TA in time is linear variation in $\mu$ and $\sigma$, and stationary $\xi$. We would expect, were larger samples used, that more complex selected models would occur. 

The table also gives optimal BMA weights $w_m$, $m=1,2,..., n_M$ estimated from the stacking procedure (Equation~\ref{Eqt:SQGY}, \cite{YaoEA18} and Section~SM9) from $n_R=100$ realisations of the LCC data-generating model only. The penultimate column of the table gives the average weight (over realisations) attributed to each fitted model complexity, together with empirical marginal 95\% uncertainty bands for each weight. {Thus on average over the $n_R=100$ sample realisations, by Equation~\ref{Eqt:SQGY}, the BMA prediction of $\Delta Q$ is dominated by contributions from the fitted LCC model (with a weight of 0.40) and the fitted ACC model (with weight 0.18). On average, the full BMA estimate for $\Delta Q$ is simply a weighted linear combination of estimates from the single model forms (CCC-QQQ), with weights as listed in column 14. It is interesting that there is large variability in fitted model weights for some fitted models. For example, the LCC weight with mean 0.40 has 95\% uncertainty band (0.00,1.00), indicating that some of the $n_R=100$ sample realisations result in weights of near zero and unity on LCC in the BMA. We attribute this large variability to the limited sample size of 86 points for model fitting, and would expect weight variability to reduce with increasing sample size. For more complex fitted models, for example QLC, QLL, QQC and QQL, the upper 97.5\%ile of the weight distribution is $\ll 1$ as might be expected.}

It is intuitively reassuring that the largest weights correspond to the LCC and ACC fits. The estimated RMSE over the $n_R$ sample realisations for BIC3-selected models is 3.06, calculated using Equation~\ref{Eqt:RMSE}; the corresponding estimate for Bayesian model averaging is 3.98, calculated using Equation~SM8. That the BIC model selection procedure outperforms the model average is perhaps not surprising here, because the choice of model selection criterion was made specifically to minimise RMSE is estimation of $\Delta Q$. Finally, the table gives BMA weights (averaged over the 5 climate ensembles) for the CMIP6 data corresponding to annual maxima from the \DSA region and the \UK GCM, noting from Figure~\ref{Fgr-Mxm-LctSA-GcmUK} that BIC3 selects the QCC model for these data. The largest weights are seen to correspond to the ACC, QLC and QCC models. It appears from these preliminary results that the stacking BMA approach favours somewhat more complex models than BIC3, perhaps more like AIC3, resulting in poorer performance in estimating $\Delta Q$. The authors plan a further study to examine the performance the stacking BMA method thoroughly.

{\color{lliw-ed}
\subsection{Uncertainty due to GCM and ensemble} \label{Sct:DscCnc:GcmEns}
%
We can summarise sources of variability in our data for changes in 100-year return values $\Delta Q$ and $\Delta^2 Q$ more precisely by estimating another statistical model. Here we estimate a linear mixed effects model (LMM, e.g. \cite{WstEA22}) using $\Delta Q$ (or $\Delta^2 Q$) as response, with climate scenarios as so-called \emph{fixed effects}, and climate models and nested climate ensemble members (for a given model) as \emph{random effects}. The advantage of this approach over linear regression (for all of scenario, model and ensemble) is that we assume the climate model and ensemble output available to us are drawn randomly from large families of models and their ensemble members. We can therefore estimate the variability in $\Delta Q$ (or $\Delta^2 Q$) which can be attributed to our random choices of climate model and their ensemble members; that is, we can estimate how much of the variability is due to (apparently) random effects from different climate models, and from different ensemble members for a given climate model. Using estimates of $\Delta \in (\Delta Q_{\text{Mxm}},\Delta Q_{\text{Mnm}}, \Delta^2 Q)$ for a given desert region, we can express the linear mixed effects model as 
\begin{eqnarray}
	\Delta_{ijk\ell} = \iota + \gamma_j + \delta_k + \zeta_{k(\ell)} + \epsilon_{ijk\ell}
\end{eqnarray}
where $\Delta_{ijk\ell} \in \mathbb{R}$ is the $i$th observation of $\Delta$ from the extreme value model output, for climate scenario $j$ ($j=1,2,3$, corresponding to scenarios \SL, \SM and \SH), for climate model $k$ ($k=1,2,...$) and ensemble member $\ell$ ($\ell=1,2,...$) of model $k$. Parameter $\iota \in \mathbb{R}$ is the intercept. Parameter $\gamma_j \in \mathbb{R}$ represents the fixed effect of scenario $j$ on response. Parameter $\delta_k \in \mathbb{R}$ represents the random effect of climate model $k$ on response, and $\zeta_{k(\ell)} \in \mathbb{R}$ the nested random effect of climate ensemble member $\ell$ for model $k$. Parameters $\epsilon_{ijk\ell} \in \mathbb{R}$ are model error terms. For fitting, we assume that each of $\delta_k$, $\zeta_{k(\ell)}$ and $\epsilon_{ijk\ell}$ are independently normally-distributed with zero means and variances $\tau_\delta^2 \ge 0$, $\tau_\zeta^2 \ge 0$ and $\tau_\epsilon^2 \ge 0$. The objective of the analysis is to estimate the set $\{\gamma_j\}_{j=1}^3$ of fixed effects, the variances $\tau_\delta^2$ and $\tau_\zeta^2$ of random effects, and the error variance $\tau_\epsilon^2$. Inference was performed using algorithms from the Julia programming language, with scenario \SL ($j=1$) as reference category for the fixed climate scenario effect. That is, to avoid issues of multicollinearity, we actually undertake estimation of the sum $\iota + \gamma_1$, together with that of scenario difference effects $\gamma_2-\gamma_1$ and $\gamma_3-\gamma_1$. Results are summarised in Table~\ref{Tbl:LME}.
}
\begin{table}[!ht]
	\centering
	\color{lliw-ed} 
		\begin{tabular}{|c|c|c|c|c|r|r|r|r|r|r|r|}
			\hline
			\multicolumn{1}{|c|}{\multirow{2}{*}{Response, $\Delta$\Tstrut}} &
			\multicolumn{1}{|c|}{\multirow{2}{*}{Region\Tstrut}} &
			\multicolumn{3}{|c|}{SSP effect\Tstrut} &
			\multicolumn{5}{|c|}{Model standard deviations\Tstrut} &
			\multicolumn{2}{|c|}{$R^2$\Tstrut} \\ \cmidrule(lr){3-12}
			& & \(\iota + \gamma_1\) & \(\gamma_2 - \gamma_1\) & \(\gamma_3 - \gamma_1\) & \(\tau_\Delta\) & \(\tau_{\text{FE}}\) & \(\tau_{\epsilon}\) & \(\tau_{\delta}\) & \(\tau_{\zeta}\) & \(R^2_{\text{FE}}\) & \(R^2_{\text{ME}}\) \Tstrut \\ \hline
			\multirow{6}{*}{\begin{tabular}{c} $\Delta Q^{\text{Mxm}}$  \\ $[\text{K}]$  \end{tabular}} & AN\Tstrut & 1.07 & 0.71 & 2.54 & 1.45 & 0.98 & 0.86 & 0.36 & 0.31 & 0.54 & 0.65 \\
			\cmidrule(lr){2-12}
			& DA\Tstrut & 0.42 & 2.75 & 8.91 & 4.04 & 1.56 & 1.29 & 0.58 & 0.65 & 0.85 & 0.90 \\
			& MO\Tstrut & 1.55 & 1.62 & 4.67 & 2.39 & 1.40 & 0.93 & 0.80 & 0.59 & 0.66 & 0.85 \\
			& SA\Tstrut & -0.31 & 3.23 & 10.61 & 4.58 & 1.13 & 0.78 & 0.69 & 0.38 & 0.94 & 0.97 \\
			& SI\Tstrut & 1.23 & 2.38 & 7.05 & 3.28 & 1.49 & 1.30 & 0.60 & 0.36 & 0.80 & 0.84 \\
			\cmidrule(lr){2-12}
			& UK\Tstrut & 1.96 & 1.64 & 5.65 & 3.31 & 2.30 & 1.37 & 1.55 & 0.83 & 0.52 & 0.83 \\
			\hline
			\multirow{6}{*}{\begin{tabular}{c} $\Delta Q^{\text{Mnm}}$  \\ $[\text{K}]$  \end{tabular}} & AN\Tstrut & 2.08 & 1.54 & 4.62 & 2.61 & 1.77 & 1.30 & 1.02 & 0.51 & 0.54 & 0.75 \\
			\cmidrule(lr){2-12}
			& DA\Tstrut & 1.36 & 1.45 & 3.81 & 3.08 & 2.65 & 1.73 & 1.89 & 0.55 & 0.26 & 0.69 \\
			& MO\Tstrut & 3.40 & 1.68 & 4.64 & 3.86 & 3.35 & 1.37 & 2.46 & 1.59 & 0.25 & 0.87 \\
			& SA\Tstrut & 1.46 & 1.14 & 3.57 & 2.22 & 1.65 & 0.75 & 1.35 & 0.53 & 0.45 & 0.89 \\
			& SI\Tstrut & 1.43 & 1.43 & 4.51 & 2.29 & 1.31 & 1.16 & 0.48 & 0.36 & 0.68 & 0.75 \\
			\cmidrule(lr){2-12}
			& UK\Tstrut & 0.93 & 0.71 & 3.38 & 2.98 & 2.60 & 1.10 & 1.94 & 1.40 & 0.24 & 0.86 \\
			\hline
			\multirow{6}{*}{\begin{tabular}{c} $\Delta^2 Q$ \\ $[\text{K}]$  \end{tabular}} & AN\Tstrut & -0.63 & -1.04 & -2.98 & 2.32 & 1.97 & 1.57 & 1.01 & 0.55 & 0.28 & 0.54 \\
			\cmidrule(lr){2-12}
			& DA\Tstrut & 0.37 & 0.32 & 1.67 & 2.71 & 2.61 & 2.48 & 0.52 & 0.65 & 0.07 & 0.16 \\
			& MO\Tstrut & -1.28 & -0.26 & -0.54 & 3.07 & 3.07 & 1.78 & 2.17 & 0.93 & 0.01 & 0.66 \\
			& SA\Tstrut & 0.30 & 1.00 & 3.05 & 2.41 & 2.05 & 1.86 & 0.74 & 0.48 & 0.28 & 0.41 \\
			& SI\Tstrut & 0.29 & 0.76 & 1.59 & 1.52 & 1.38 & 1.33 & 0.16 & 0.30 & 0.18 & 0.23 \\
			\cmidrule(lr){2-12}
			& UK\Tstrut & 1.41 & 0.79 & 1.98 & 2.34 & 2.19 & 1.71 & 1.15 & 0.92 & 0.12 & 0.46 \\
			\hline
		\end{tabular}%
	\caption{{Summary of linear mixed effects modelling for the changes in reponses $\Delta \in (\Delta Q^{\text{Mxm}}, \Delta Q^{\text{Mnm}}, \Delta^2 Q)$. Columns under ``SSP effect'' show intercept $\iota$ and fixed effect parameter estimates $\gamma_j$ for the change in 100-year return value of the given variable over the period (2025,2125) under scenario $j$, unit: K. Columns under ``Model standard deviations'' provide estimates of the various standard deviations of model fitting, unit: K. $\tau_\Delta$: the (full unconditional) standard deviation of the response $\Delta$; $\tau_{\text{FE}}$: the model error standard deviation after fitting only the fixed effects (FE, of climate scenario); $\tau_{\epsilon}$: the model error standard deviation after fitting the full mixed effects model. $\tau_{\delta}$: standard deviation of climate model random effect; $\tau_{\zeta}$: standard deviation due to nested random effect of climate ensemble within climate model. $R^2$ statistics are also provided, from fitting only the fixed effects (FE), and from fitting the full mixed effects (ME) model.}}
\label{Tbl:LME}
\end{table}

{Inferences regarding the sizes of fixed effects in Table~\ref{Tbl:LME} for all of $\Delta Q^{\text{Mxm}}$, $\Delta Q^{\text{Mnm}}$ and $\Delta^2 Q$ across regions are similar to the means estimated in Table~\ref{Tbl:ExpPrb}, recalling that the mixed-effects model is parameterised in terms of differences $\gamma_j-\gamma_1$, $j=2,3$. The $R^2$ estimates indicate how much of the total response variance is explained by the fixed effects alone, and by the combination of fixed and random effects in the mixed effects model. Thus we see in general that the fixed effects are more informative for $\Delta Q^{\text{Mxm}}$ than for $\Delta Q^{\text{Mnm}}$ and $\Delta^2 Q$. However, in general, we see that the mixed effects model explains variation in both $\Delta Q^{\text{Mxm}}$ and $\Delta Q^{\text{Mnm}}$ equally well. That is, the variation in these responses is well characterised by a random effects model for climate model and (nested) climate ensemble. Estimates for $\Delta^2 Q$ are less convincing (indicated by lower $R^2$ values). By definition, comparison of values of $\tau_\Delta$, $\tau_{FE}$ and $\tau_\epsilon$ would yield similar findings to comparison of $R^2$ values. Comparison of $\tau_\delta$ and $\tau_\zeta$ allows quantification of the relative sizes of sources of uncertainty due to GCM and scenario given GCM. We see, for example, for $\Delta Q^{\text{Mnm}}$ in the \DSA region, that $\tau_\delta$ (=1.35) is more than twice the size of $\tau_\zeta$ (0.53), indicating that the between-GCM variation is considerably larger than the between-\ed{ensemble} variation given GCM. In contrast, for $\Delta Q^{\text{Mxm}}$ in the \DDA region, $\tau_\delta$ (=0.58) is slightly smaller than $\tau_\zeta$ (0.65), indicating similar contributions from between-GCM variation and between-\ed{ensemble} variation given GCM. Overall, for $\Delta Q^{\text{Mxm}}$ and $\Delta Q^{\text{Mnm}}$, we conclude that between-GCM variation is considerably larger than between-\ed{ensemble} variation given GCM.}

\subsection{Conclusions} \label{Sct:DscCnc:Cnc}

{Changes in the distribution of extreme temperatures in desert regions due to climate change are likely to present severe societal challenges. For example, \cite{NtmEA26} estimates future extreme temperatures in the Middle East-North Africa (MENA) region, including the Sahara, predicting (under 2$^\circ$C global warming) that the return period for extreme 50-year heat events (in today's climate) will reduce to just a few years. Under 3$^\circ$C global warming, half of the population of the MENA region will be exposed to dangerous heatwaves (and see e.g. \cite{YzlEA23}). The decadal report of the Scientific Committee for Antarctic Research (SCAR; \cite{ChwEA2022}) argues drivers of global climate change are also impacting the Antarctic and Southern Ocean environments, and that the most significant potential impact of Antarctic changes will be via global mean sea level rise and its consequences for coastal regions globally. They also anticipate global influence on extreme climate and weather events, droughts, wildfires and floods, and ocean acidification. Estimating the extent of these changes from data using statistical models is not straightforward.}

When competing models of different complexity are considered in non-stationary extreme value modelling using small samples, care must be taken in model selection. The relative performance of different ``information criteria'' (including AIC, BIC, DIC and WAIC) for model selection with respect to a specific prediction task cannot be easily anticipated. We recommend that a simulation study is undertaken, using relevant data, to identify which information criteria perform best for the prediction task. For the current work, focussing on predicting the difference $\Delta Q$ in the 100-year return value over the period (2015,2125), of regional annual maxima and minima of \TA using Bayesian inference, we find that the Bayesian information criterion (BIC) provides optimal predictive performance. We also demonstrate that Bayesian model averaging provides a feasible alternative to model selection, but - using the stacking approach considered here - at considerably increased computational cost.

{Using optimal model selection, for regional annual maxima and minima of \TA in hot and cold desert regions (and the temperate \DUK ``control'' region), $\Delta Q$ increases with scenario forcing, for most combinations of regions, climate models and ensembles, for both regional annual maxima and minima. Largest increases for $\Delta Q^{\text{Mxm}}$ are seen in the \DDA, \DSA, \DSI and \DUK regions under the \SH scenario.  Changes in $\Delta Q^{\text{Mnm}}$ are more similar across regions than changes in $\Delta Q^{\text{Mxm}}$. Using inferences aggregated over GCM and climate ensemble, there is evidence for significant increases in $\Delta Q$ for regional annual maxima under scenarios \SM and \SH for all regions considered. Weaker increasing trends (with larger relative uncertainties) are observed for regional annual minima. For \DDA, \DSA, \DSI and \DUK, there is evidence for significant asymmetric warming (with $\Delta^2 Q>0$): 100-year return values for regional annual maxima are increasing more quickly than those for regional annual minima under \SH. For the \DAN region, there is evidence for significant asymmetric warming (with $\Delta^2 Q<0$) under both \SM and \SH. Differences in trends for $\Delta Q$ and $\Delta^2 Q$ across desert regions reflect the different physical processes responsible for extremes of temperature in those regions, as outlined in Section~\ref{Sct:Int}. The general increase observed in $\Delta Q$ across all desert regions is an early indicator for changes in the global climate. }

{In this work, we use the RMSE of $\Delta Q$ in out-of-sample prediction as the scoring rule with respect to which the information criterion providing optimal predictive performance can be identified. Of course, were we to adopt a different scoring rule for model evaluation, we might identify a different optimal information criterion. Moreover, the choice of optimal information criterion also depends on the specific application under consideration. Since the simulation methodology is sensitive to both the scoring rule and the application data, it provides a general-purpose scheme to identify the optimal information criterion, for any combination of scoring rule and application of interest.}

{Model selection in empirical inference is a source of uncertainty, since the sample data will not always be sufficiently informative to allow correct identification of model form. Nevertheless, it is reasonable to expect that a carefully selected model is likely to perform better than an arbitrarily chosen model form. The appeal of Bayesian model averaging (BMA) is that is offers a means to avoid the problem of model selection, instead providing a predictive model as a weighted sum of potential model forms. Model selection is generally computationally a more straightforward task compared with BMA (see e.g. SM9). Despite its intuitive appeal, the results in Table~\ref{Tbl:BMA} illustrate the general point that there is no reason to expect BMA to out-perform a carefully chosen single model form in practice.}

{In practice, characterising future extreme temperatures for the Earth's arid regions in isolation is unlikely to be sufficient in understanding climate change impact and preparing adequate mitigation in those regions. A more comprehensive assessment of compound extremes of temperature in conjunction with e.g. precipitation and related hydrological extremes is required (e.g. \cite{BcvEA23}, \cite{FtsEA26}), incorporating quantification of the spatio-temporal evolution and dependence of these extremes.}

{It would be interesting to consider quantifying the evolution over time of the full distribution of regional annual temperatures (including minima and maxima) and suitably-chosen intermediate distributional quantiles using a multiple non-crossing quantile regression (e.g. \cite{DaiEA23}) to complement the GEV regressions presented here.}

\bibliography{C:/Philip/Git/Cod/LaTeX/phil}

\end{document}